\documentclass[3p,11pt,times]{elsarticle}
\usepackage{ecrc}
\volume{31}
\firstpage{1425}
\journalname{Applied Mathematical Modelling}
\runauth{Andonowati, N. Karjanto and E. van Groesen}
\jnltitlelogo{\large  Applied Mathematical Modelling}
\usepackage{amssymb}
\usepackage[figuresright]{rotating}

\usepackage[all]{nowidow}
\usepackage{subfigure}
\usepackage{booktabs}
\usepackage{tabularx}
\newcolumntype{R}{>{\raggedleft\arraybackslash}X}
\newcolumntype{L}{>{\raggedright\arraybackslash}X}
\newcolumntype{C}{>{\centering\arraybackslash}X}

\biboptions{numbers,sort&compress}
\usepackage{fancyhdr}
\usepackage{psfig}
\usepackage{graphics}
\usepackage{graphicx}
\usepackage{epsfig}
\usepackage{url}     
\usepackage{amsmath}  

\begin{document}

\begin{frontmatter}
\dochead{}
\title{Extreme wave phenomena in down-stream running modulated waves}
\author[0,1]{Andonowati\corref{cor1}}
\ead{aantrav@attglobal.net}
\cortext[cor1]{Corresponding author.}
\author[1]{N. Karjanto\,}
\author[1]{E. van Groesen\,}
\address[0]{\,Department of Mathematics, Institut Teknologi Bandung, Jl. Ganesha 10 Bandung, 40123, Indonesia}
\address[1]{\,Department of Applied Mathematics, University of Twente, The Netherlands \\
{\hfill} \\
{\upshape Received 1 June 2005; received in revised form 1 March 2006; accepted 12 April 2006}\\
{\upshape Available online 27 June 2006}}

\begin{abstract}
Modulational, Benjamin-Feir, instability is studied for the down-stream evolution of surface gravity waves. An explicit solution, the soliton on finite background, of the NLS equation in physical space is used to study various phenomena in detail. It is shown that for sufficiently long modulation lengths, at a unique position where the largest waves appear, phase singularities are present in the time signal. These singularities are related to wave dislocations and lead to a discrimination between successive `extreme' waves and much smaller intermittent waves. Energy flow in opposite directions through successive dislocations at which
waves merge and split, causes the large amplitude difference. The envelope of the time signal at that point is shown to have a simple phase plane representation, and will be described by a symmetry breaking unfolding of the steady state solutions of NLS. The results are used together with the maximal temporal amplitude
MTA, to design a strategy for the generation of extreme (freak, rogue) waves in hydrodynamic laboratories.\\
{\footnotesize \copyright \ 2006 Elsevier Inc. All rights reserved.}\\
\begin{keyword}
Extreme waves \sep Freak waves \sep Rogue waves \sep Phase singularity \sep Wave dislocation \sep Benjamin-Feir instability \sep 
Maximal temporal amplitude \sep Soliton on finite background \sep Wave generation \sep Hydrodynamic laboratory \\
{\hfill}\\
\MSC[2008] 35Q51 \sep 76B25 \sep 35Q55 \sep 35B30 \sep 76E30 \sep 76B15
\end{keyword}
\end{abstract}
\end{frontmatter}

\section{Introduction}

Weakly nonlinear, dispersive evolution equations describe phenomena in such areas as nonlinear optics and gravity-driven surface waves on a layer of fluid. When the diverging effect of dispersion and the converging effect of nonlinearity balance each other, phenomena like solitons and soliton-like interactions appear for exponentially confined waves, their periodic equivalents as well as for the collective behaviour of waves in wave groups \cite{AkhAnk,Whitham}. Amplitude increase and large deformation of wave profiles are characteristic and relatively easy observable phenomena. However, the details of the processes and the interaction between them remain rather difficult, despite many investigations. In fact, interesting phenomena can be studied in detail theoretically only in simplified models. Fortunately,
the important phenomena seem to be generic enough to be captured by these models, so that what we learn from these models is also useful to describe important
aspects of more realistic situations. This paper will contribute to this type of research in the following way.

Taking the well-known model for the collective behaviour of wave groups that is called the NLS-(NonLinear Schr\"{o}dinger) equation, we will study in much detail one specific solution. This solution, the so-called soliton of finite background (SFB), can be written down explicitly, and we will extract from it detailed information about such processes like (extreme) amplitude amplification, phase singularity and wave dislocation, a generic form of extreme waves, and spectral evolution and interaction of wave modes. Some of the properties to be described in this paper will be known at least qualitatively, but to our knowledge the explicit description of the properties given below in the physical setting leads to some surprising and illuminating results, in particular about the wave dislocation and properties of the extreme signal.

The direct motivation of this paper stems from the need of hydrodynamic laboratories to generate `extreme' waves, often also called `freak' or `rogue' waves. In such waves, ships and marine constructions can be tested under extreme sea-conditions. Technical restrictions on the wavemaker make it impossible to generate such large waves directly; the nonlinearity in the physical phenomenon when the wave is running downstream is exploited to obtain the large waves. Hence, to generate large waves at a certain position in a long wave tank, the motion of a wave maker at one end of the wave tank will generate moderately small waves that start to travel downstream
along the tank over an initially still water level. We can view this evolution as a \textit{signalling problem}: the wave elevation at each position defines a time-signal, so that the down-stream evolution of waves correspond to the subsequent signals as a response to the time signal at the position of the wave maker. Due to the non-linear effects determined by nature, large deformations may appear and large and steep waves can emerge within the wave tank. Such strong non-linear effects and the appearance of large amplitude amplification in generated waves can be seen in experimental results \cite{JHGrRene01,JHGrRene02} as well as numerical results \cite{GrJH01,JH,JHAan,JH02}. It is of the laboratory's interest to generate extreme waves in a deterministic way based on reliable theoretical predictions, for which we will use the SFB solution of the NLS model in this paper.

Different from the deterministic wave generation in a laboratory, in real seas, many investigations on extreme (freak) waves in random sea states have been carried out in the past years. Then an extreme wave is understood \cite{Dean} as a large wave with wave height $H$ exceeding the significant wave height $H_{s}$ by a factor of $2.2$. Most of these studies are for initial value problems: different from the signalling problem to be considered here, then the temporal evolution of an initial wave elevation over a spatial domain is investigated.

\begin{figure}[h]
\begin{center}
\includegraphics[width=0.85\textwidth]{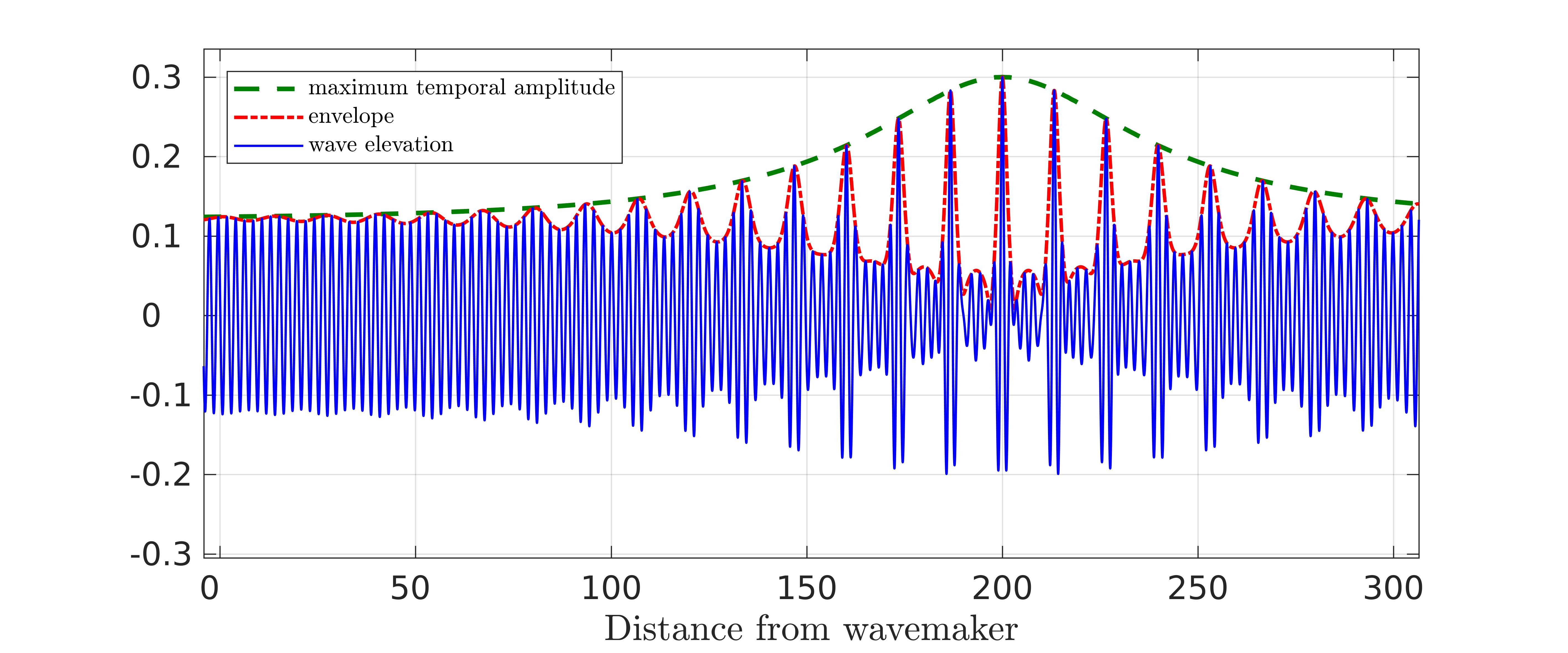}
\end{center}
\vspace*{-0.5cm}
\caption{A snapshot of a wavefield of waves generated at the left, while travelling to the right; the physical unit along the axis is in meter, and the wavelength is approximately equal to the depth of the layer. The dash-dotted line denotes the envelope of the wave groups, which consist of about six waves, while the dashed line denotes the MTA, the maximum temporal amplitude, which is the largest possible wave elevation at a fixed position. See the text for a more extensive description.}
\label{spatWavefieldMTA}
\end{figure}

In all such investigations, extreme waves, often occurring in wave group structures, are of most interest and discussed in several papers, such as \cite{Donelan,GroAanKar04Brest,Longuet,Osb00,Phillips}. The nonlinear self-focusing phenomenon is described to result in very steep waves of high amplitude that arise intermittently within wave group structures and that govern the dynamics of extreme wave occurrences  \cite{Dysthe,Hen99}. Modulational instability of Benjamin-Feir type \cite{BenjaminFeir} is often described as the cause of self-focusing that leads to the generation of extreme waves. In this paper, we will investigate this effect in more detail. To facilitate the understanding of the phenomenon, we refer to Fig.~\ref{spatWavefieldMTA}.

This figure shows, in real dimensions, at a certain time a spatial view of (first order) waves with a wavelength of approximately the depth of the layer of fluid, about $5$ m. The waves are generated at the left, at $x=0$ in the figure, and travel to the right. A number of waves (about six waves in the figure) of varying amplitude, form a wave group, which has the envelope as depicted by the dashed line in the figure. These wave groups are seen to change substantially depending on
the distance from the wavemaker: from small modulations close to the wave maker, to severely amplified and distorted wave groups near the point where the waves of maximal amplitude appear (in the figure at about $200$ m), after which the groups decrease again for increasing distance. This plot of the waves will change at each moment, the waves translate with the phase velocity and the envelope with the group velocity. The maximal wave height at a certain position is given by the dotted line, which denotes the maximal temporal amplitude MTA. The MTA, introduced originally in \cite{AB}, is a useful concept in the study of the long-time behaviour of wavefields; see also \cite{AanRnBr,AanRnBrJPO} where it was used to study the phenomenon of wave breaking.

Fig.~\ref{spatWavefieldMTA} shows wave groups that will be studied in this paper. These waves evolve from a modulated monochromatic wave. In fact, we will use a fully nonlinear extension of the linear modulational (Benjamin-Feir) instability of a monochromatic wave which is described by a specific explicit solution of NLS, called the soliton on finite background (SFB) mentioned above. When translated to physical variables, the SFB shows additional phenomena compared to its description as
an amplitude in the NLS-setting. Actually, the SFB is not an isolated solution, but is embedded in a family of SFB-solutions. This family depends on three essential parameters: the frequency $\omega$ of the monochromatic `carrier' wave, the modulation-length $\nu$, which together with the frequency $\omega$ determines the number of waves in one wave group, and the maximal amplitude which is related to the asymptotic value of the monochromatic wave amplitude. SFB solutions are described in detail in \cite{AkhAnk} and have already been considered in \cite{Osb00,Osb01} as a possible description of large amplitude increase of surface waves leading to `freak', or `rogue' waves. In this paper, we will add substantially to these previous investigations.

Actually, besides SFB, large amplitude phenomena are also present in two different but closely related explicit solutions of NLS, the class of Ma-solutions and the single polynomial solution of Peregrine; see \cite{Akhmediev87,Dysthe99,Karjanto04,Ma79,Peregrine83} and \cite{GroAanKar06DPA} for a unified description. The last solution appears for both SFB and Ma-solutions as a limiting case and has polynomial decay in time and space to a uniform wave train. Since our interest is in the signalling problem with a slowly modulated wave at the initial position, i.e. the situation of the onset of the BF-instability at one specific position (the wave maker), it leads us to select the SFB-solution for detailed consideration. Large amplitude amplification of monochromatic wave trains is also observed to occur in water wave models different than the NLS model. For example, in Green-Naghdi Level III equations, the band-modulated Stokes waves show amplitude amplifications up to
a factor of $3.1$ \cite{Ert00}. Numerical investigations of modulated wave trains within a fully nonlinear irrotational numerical flow solver show that the amplitude amplification can be even larger than $3.3$ for a wide range of initial wave steepness \cite{Hen99}; a similar result is demonstrated also in \cite{Zak02}.

We now briefly summarize the main contributions reported in this paper. We will describe that when the modulation is long enough, at the extreme position
(where the wave heights are largest), phase singularities appear periodically in time. These phase singularities correspond to wave dislocations of two
types, dislocations of wave splitting and of wave merging, at which energy flow in opposite direction takes place. This leads to the understanding that these
successive splitting and merging phenomena are responsible for a temporal separation and discrimination between `extreme' waves and intermittent waves of much smaller amplitude. At the extreme position, this amplitude difference can be essentially larger than when the maximal amplitude is compared to the amplitude of the far-field; this last amplification factor is at most $3$ \cite{KarGroPet02}. We will show, in fact, that the time signal at the extreme position has very special properties, and can be described by a Newton type of second order equation. The (classical-mechanical) potential of that system is a symmetry-breaking unfolding of the potential that describes the steady states of NLS, such as the standard soliton and periodic oscillations in the amplitude around an equilibrium value of a nonlinear monochromatic
mode. In the limit for long modulation lengths, the extreme signal tends to a soliton on finite, \textit{negative}, background. The extreme waves in this signal are almost independent of the number of intermittent waves, which number varies from zero to infinity. This shows that the extreme waves have, practically speaking, a generic form. Aside from clearly observable phenomena as wave dislocations and the generic extreme waves, the SFB solutions show interesting properties that are less easy to observe directly. This includes the spatial evolution of the spectrum which shows large energy exchange between modes as a consequence of different contributions from the extreme and intermittent waves.

In this paper, we address these properties and discuss the applicability for the sake of deterministic extreme wave generation. Actually, the findings here have motivated and designed a series of experiments in the large wave tank ($200$ m long, $4$ m deep) of MARIN, the Maritime Research Institute Netherlands \cite{expMARIN04}. These experiments revealed that most of the generated waves did not `break', which could not have been predicted within the NLS-model that is used here; numerical indications that breaking will occur only in the most extreme cases have already been reported~\cite{AanRnBrJPO}.

The organization of this paper is as follows. In the next Section~\ref{spatial}, the NLS equation and the explicit family of SFB solutions are presented. 
The phenomenon of phase singularity and wave dislocation, and the explanation of the energy exchange, is described in Section~\ref{ps}. In Section~\ref{es}, the extreme signal of SFB will be investigated in detail. In Section~\ref{dsse} the energy exchange during down-stream evolution is shown in spectral components of the successive time signals. Various cases for characteristic parameters and samples of extreme wave generation in laboratory coordinates are presented in Section~\ref{dseg}, together with the further explanation of the maximal temporal amplitude. In the last section, we give a brief summary and some concluding remarks.

\section{Spatial NLS and soliton on finite background} \label{spatial}

\subsection{The NLS equation} \vspace*{0.25cm}

In the following, we use the notation $\eta(x,t)$ for the physical wave amplitude. The variables $\eta$, $x$, and $t$ are non-dimensionalized and are related to laboratory variables by the scaling $\eta_{lab} = \eta \, \ast h$, $x_{lab} = x \ast h$ and $t_{lab} = t \ast \sqrt{h/g}$ where $h$ is water depth and $g$ is gravitational acceleration. For linear dispersive wave equations, a simple linear monochromatic wave is described by $\eta(x,t) = A \exp \left[i(k_{0} x - \omega_{0} t) \right] + \text{cc}$, where $\omega_{0}$ is the frequency and $k_{0}$ is the wavenumber that are related by the dispersion relation, say $\omega_{0} = \Omega\left(k_{0} \right)$, $A$ is a constant amplitude and `cc' denotes complex conjugation of the previous term. For small amplitude surface waves, the linear description is governed by the dispersion given by $\Omega \left(k \right) = \sqrt{k\tanh(k)}$.

For weakly nonlinear phenomena, the deformation of a linear monochromatic wave is commonly described by a series expansion in the (small) amplitude. The resulting wave group centred around the basic frequency $\omega_{0}$ then describes the spatial deformation with a complex amplitude $A$ that is allowed to vary in time and space, i.e. in the lowest order
\begin{equation}
\eta(x,t) \approx A(\xi,\tau) \exp\left[i(k_{0}x - \omega_{0}t) \right] + \text{cc}. \label{physNLSrelation}
\end{equation}
Here local coordinates in a coordinate system moving with the group velocity $V_{0} = \Omega^{\prime}(k_{0})$ are used: $\tau = t - x/V_{0}$, $\xi = x$. For weakly
nonlinear physical phenomena, the complex amplitude $A(\xi,\tau)$ satisfies a spatial NLS-type of equation for solutions for which there is a balance between dispersive and nonlinear effects:
\[
\partial_{\xi} A + i \beta\partial_{\tau}^{2}A + i \gamma|A|^{2}A = 0.
\]
The parameters $\beta$ and $\gamma$ depend on the monochromatic frequency; with a KdV-type of equation to model the surface waves, these parameters are easily written down as follows (see \cite{Gro98}) $\beta = -\frac{1}{2}\Omega^{\prime\prime}(k_{0})/V_{0}^{3}$, $\gamma = \frac{9}{4}k_{0}(\sigma_{0} + \sigma_{2})/V_{0}$, with $\sigma_{0} = 1/(\Omega^{\prime}(k_{0}) - \Omega^{\prime}(0))$, $\sigma_{2} = k_{0}/(2\Omega(k_{0}) - \Omega(2k_{0}))$ and $V_{0}$ is the group velocity. We will restrict to the case of interest in the `self-focussing' regime for which sign$(\beta \gamma) > 0$, which is a restriction on the maximum wavelength of the carrier waves: $k_{0} > k_\text{crit}$ with $k_\text{crit} \approx 1.23$. 

\noindent
\textbf{Remark. \;} In all of the following, we will use relation (\ref{physNLSrelation}) to translate properties of the NLS equation to the physical variables, and
thereby do not take higher order contributions into account. This is done for ease of presentation. But it should be noted that, for instance, the characteristic Stokes phenomenon from second-order terms is therefore missing, and that the quantitative wave heights are not very accurate. Indeed, for laboratory practice, 
second-order effects have to be added, which may increase the wave height with up to~10\%.

\subsection{The family of SFB solutions} \vspace*{0.25cm}

A so-called nonlinear monochromatic wave of constant amplitude $r_{0}$ is described by
\[
A_{0} = r_{0}e^{-i \gamma r_{0}^{2} \xi},
\]
and corresponds to a harmonic physical wavefield with (nonlinearly) adjusted dispersion relation:
\[
\eta_{nm}=2r_{0}\cos \left((k_{0} - \gamma r_{0}^{2}) x - \omega_{0}t \right).
\]
\begin{figure}[h]
\vspace*{-0.5cm}
\begin{center}
\includegraphics[width=0.4\textwidth]{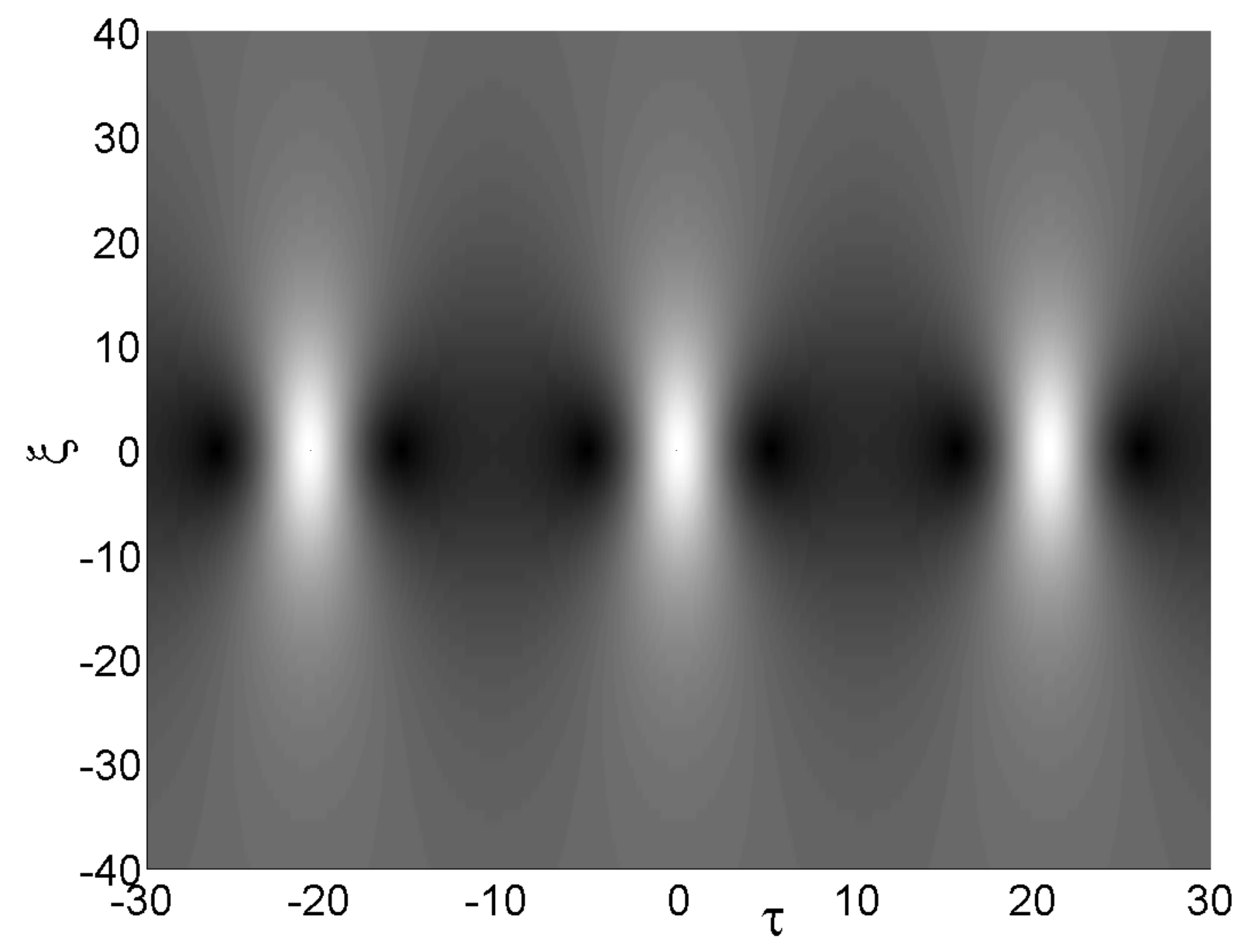} \qquad \qquad 
\includegraphics[width=0.4\textwidth]{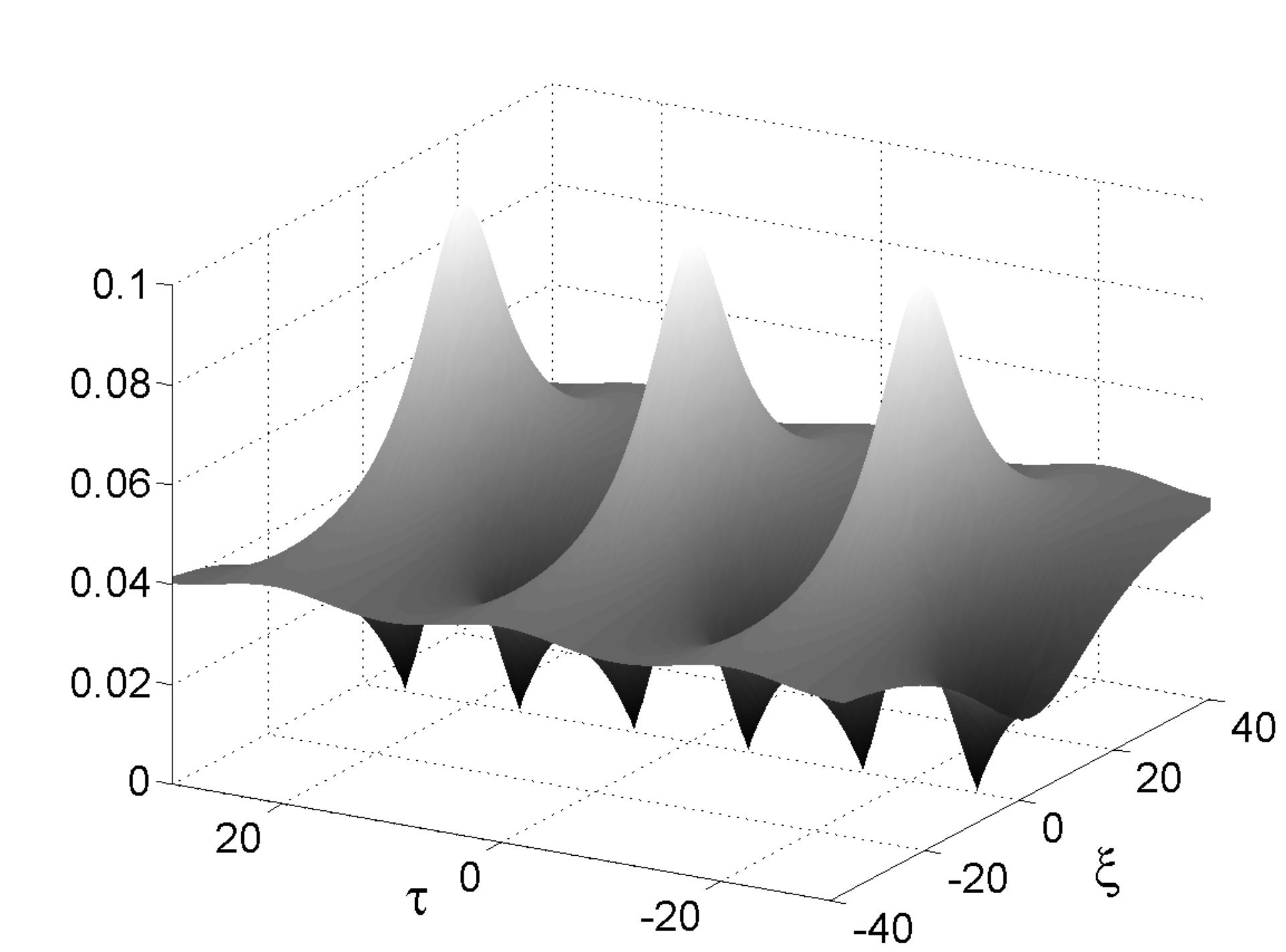} 
\end{center}
\vspace*{-0.5cm}
\caption{Density plot (left) with horizontally the time $\tau$ and vertically $\xi$, and a 3D plot (right) of $2a = 2|A|$, i.e. twice the absolute value of $A_\text{SFB}(\xi,\tau)$ as given by (\ref{explicitA}), for the set of parameter values: $\hat{\nu} = 1,\omega = 2.5$ and $M = 0.1$.} \label{2}
\end{figure}

Another well-known solution of the NLS equation is the soliton solution which decays to zero at $\xi=\pm\infty$. We will return to this `standard' soliton in Section~\ref{es} to relate it to a description of the extreme signal. The solution of main interest in this paper is the soliton on finite background, SFB. This is the solution that arises when a nonlinear monochromatic wave is perturbed with a long-period modulation of the basic frequency, and that describes the connection of a nonlinear monochromatic wave of amplitude $2r_{0}$ at $\xi = -\infty$ to a similar (shifted) wave of the same amplitude at $\xi=\infty$, while undergoing a Benjamin-Feir instability stimulated by the modulation. This linear instability leads to exponential growth of the `side-band' perturbations. In fact, perturbations in frequency at sidebands $\omega_{0} \pm \nu$ will grow (or decay) with the Benjamin-Feir growth rate $\sigma(\nu) = \beta \nu \sqrt{2\nu_{\ast}^{2} - \nu^{2}} = \gamma r_{0}^{2} \hat{\nu} \sqrt{2 - \hat{\nu}^{2}}$ provided this is real, i.e. provided $\hat{\nu} < \sqrt{2}$ where $\hat{\nu} = \nu/\nu_{\ast}$ with 
$\nu_{\ast} = r_{0} \sqrt{\gamma/\beta}$ a normalized frequency; the maximal growth rate is attained for $\hat{\nu} = 1$ and has value $\gamma r_{0}^{2}$.
The linear modulational instability is described precisely by the asymptotic behaviour of the SFB, which solution can, therefore, be seen as a continuation of the Benjamin-Feir instability into the nonlinear regime (see e.g. \cite{AkhAnk}). This SFB is explicitly given by
\begin{equation}
A_\text{SFB}(\xi,\tau;\omega_{0},r_{0},\nu)=A_{0} \frac{(\hat{\nu}^{2} - 1) \cosh(\sigma \xi) - i \hat{\nu} \sqrt{2 - \hat{\nu}^{2}} \sinh(\sigma \xi) + 
\sqrt{1 - \hat{\nu}^{2}/2} \cos(\nu \tau)}{\cosh(\sigma \xi) - \sqrt{1 - \hat{\nu}^{2}/2} \cos\nu\tau}. \label{explicitA}
\end{equation}
\begin{figure}[h]
\begin{center}
\includegraphics[width = 0.4\textwidth]{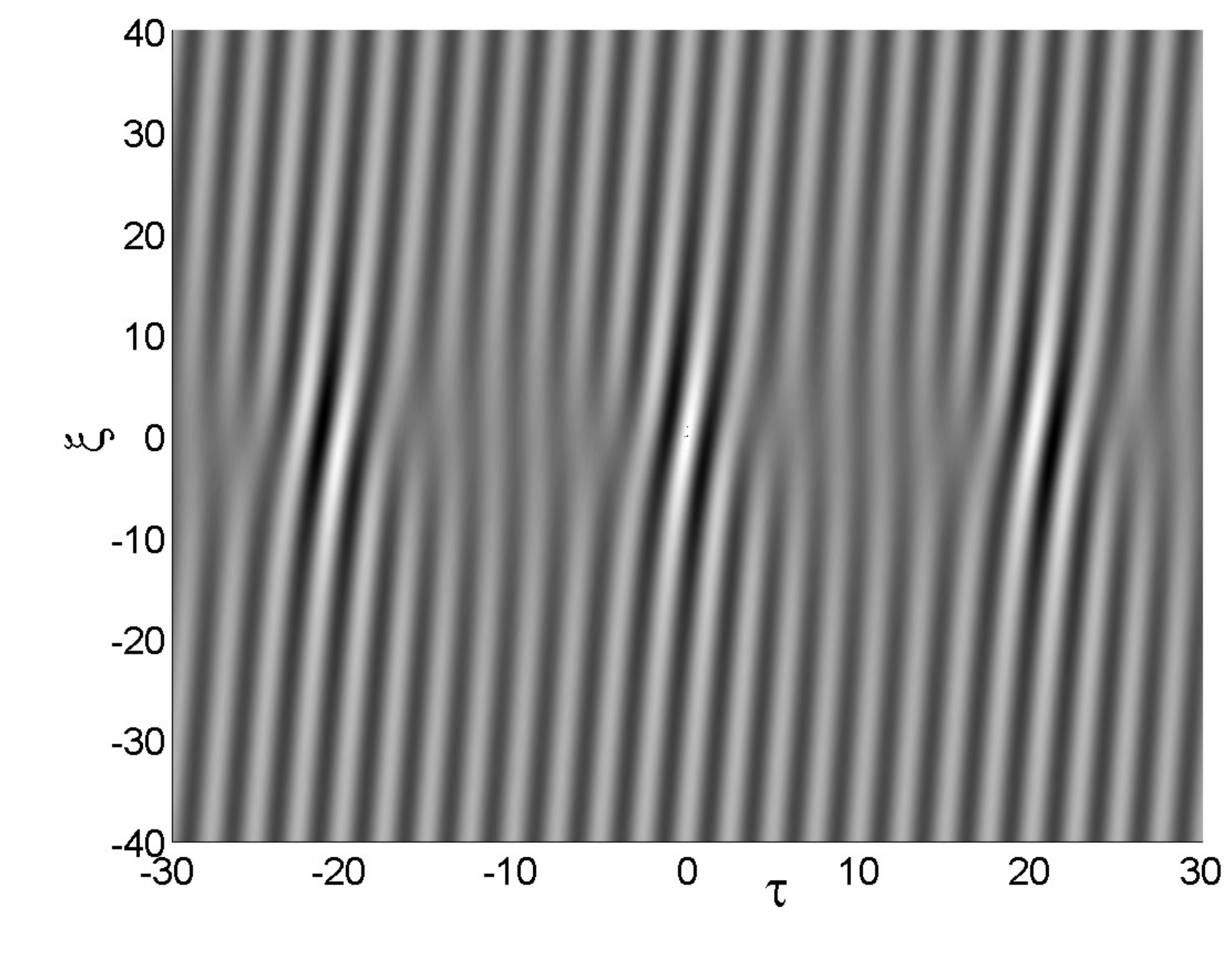} \qquad \qquad 
\includegraphics[width = 0.4\textwidth]{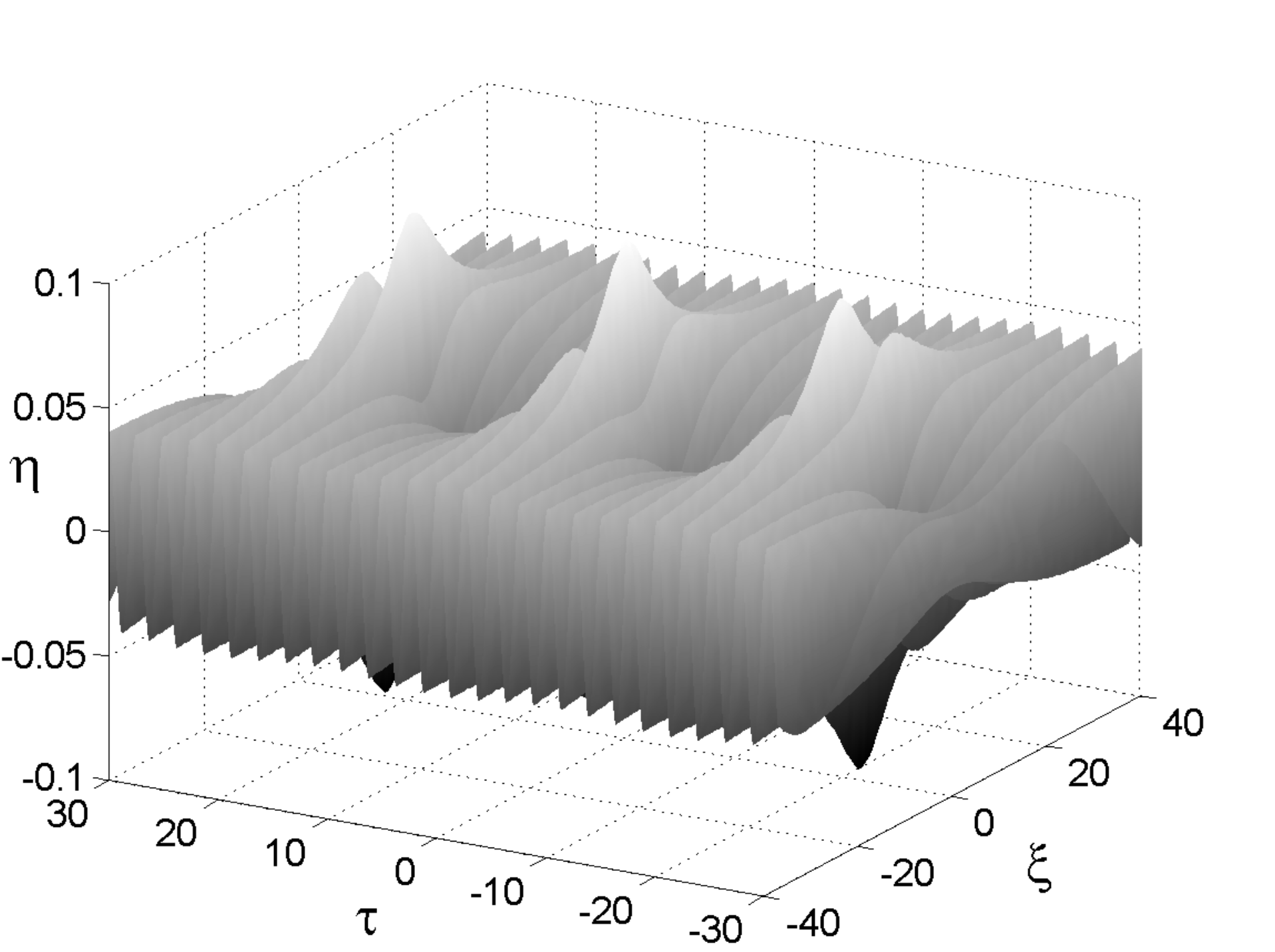}
\end{center}
\vspace*{-0.5cm}
\caption{Density plot (left), with horizontally the physical time shifted with the group velocity and vertically the physical space, and 3D plot (right) of the real wave field $\eta_\text{SFB}(x,t)$ corresponding to the case depicted in Fig.~\ref{2}.} \label{3}
\end{figure}

Note the periodicity in $\tau$ with period $T_\text{mod} = 2\pi/\nu$ and the soliton-type of shape in $\xi$ on top of the level at value $r_{0}$ with a phase shift $2\phi$ from $\xi = -\infty$ to $\xi = \infty$ with $\tan \phi = \hat{\nu} \sqrt{2 - \hat{\nu}^{2}}/(\hat{\nu}^{2}-1)$.

Through formula (\ref{physNLSrelation}) this SFB solution gives an approximate solution of the physical system; we will denote this solution in the original
space-time variables by $\eta_\text{SFB}$, depending on three parameters, $\omega_{0},r_{0}$ and $\hat{\nu}$. By writing the complex amplitude $A$ in `polar' coordinates by introducing a real amplitude $a$ and phase $\phi$ according to $A = a e^{i\phi}$, the real wavefield is given by
\begin{equation}
\eta_\text{SFB} = 2a \cos \Phi, \qquad \text{with} \quad 
\Phi = k_{0}x - \omega_{0}t + \phi(x,t). \label{physicalSFB}
\end{equation}

In order to give a graphical illustration of the contents of the explicit formula (\ref{explicitA}), we show in figure Fig.~\ref{2} a density plot and a 3D plot of the amplitude $2a = 2|A|$; the corresponding real wave field $\eta_\text{SFB}$ is shown in Fig.~\ref{3}. 
Parameters used for these plots are $\hat{\nu} = 1$, $\omega = 2.5$ and $M = 0.1$.

From the explicit formula (\ref{explicitA}) it is easily seen that the maximal amplitude of $\eta_\text{SFB}$, which will be denoted by $M$, is attained for $\xi = \tau = 0$, and has value given by
\begin{equation}
M = 2 r_{0} \left( 1 + 2\sqrt{1 - \hat{\nu}^{2}/2} \right). \label{rel M and r0}
\end{equation}

\begin{figure}[h]
\begin{center}
\includegraphics[width = 0.4\textwidth]{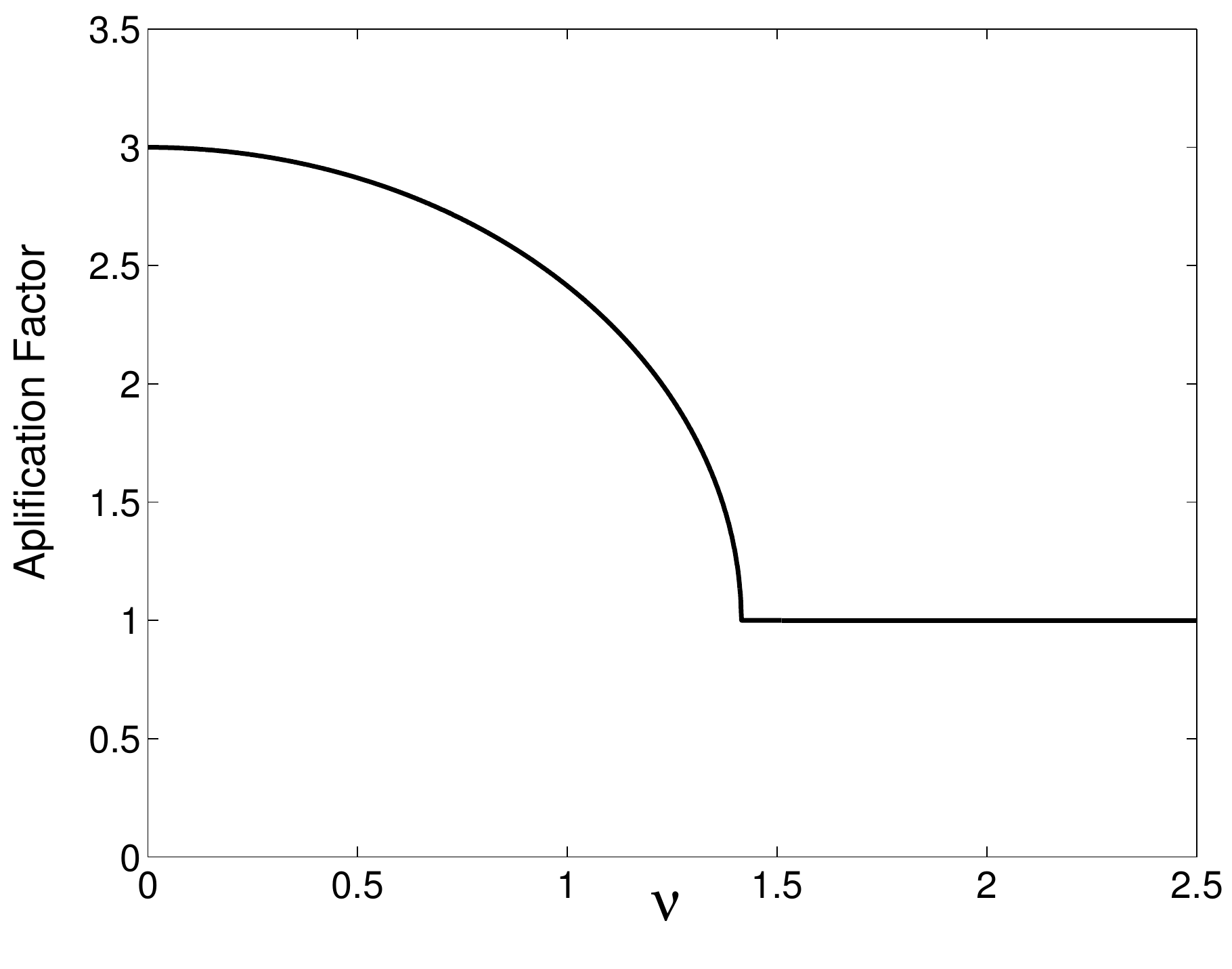} 
\end{center}
\vspace*{-0.7cm}
\caption{The plot of the maximum amplification factor (ratio of the maximal wave amplitude and the amplitude of the linear monochromatic wave at infinity) as a function of $\hat{\nu}$, with limiting value $3$ for infinitely long perturbations. Values larger than $1$ are only found within the Benjamin-Feir instability interval $\hat{\nu} \in (0,\sqrt{2})$.} \label{4}
\end{figure}

This expression shows that the \textit{maximum amplitude amplification}, defined as the ratio of the maximal amplitude and the amplitude of the linear monochromatic wave at infinity $\alpha = M/(2r_{0})$, is monotonically increasing for decreasing $\hat{\nu}$ with limiting value $3$ attained for $\hat{\nu} \rightarrow 0$, see Fig.~\ref{4}.

In the next sections, we will further investigate several physical phenomena that are hidden in the explicit expressions given above, and exploit the formulas to predict and prescribe extreme wave generation in hydrodynamic laboratories.

\section{Phase singularity} \label{ps}

\subsection{Phase jumps for indefinite envelopes} \vspace*{0.25cm}

Investigating the complex amplitude $A_\text{SFB}$, it is noted that this amplitude is real-valued (only) at $\xi = 0$, which is the extreme position where the
maximal amplitude is obtained for $\tau = 0{\,\text{mod}} \left(T_{\text{mod}} \right)$, where $T_{\text{mod}} = 2\pi/\nu$ is the modulation period. 
The signal of the physical solution $\eta_\text{SFB}$ at the extreme position $\xi = x = 0$ will be called the \textit{extreme signal}. This extreme signal is given by $\eta_\text{SFB}(x = 0, \tau; \omega_{0}, r_{0}, \nu) = s_{0} \left(\tau \right)$, with
\begin{equation}
s_{0}(\tau) = 2S(\tau) \cdot \cos(\omega_{0}\tau) \qquad \text{with} \quad 
S(\tau) = r_{0} \cdot\frac{\hat{\nu}^{2} - \left[1 - \sqrt{1-\hat{\nu}^{2}/2}
\cos(\nu \tau)\right]}{1 - \sqrt{1 - \hat{\nu}^{2}/2}\cos\nu\tau},  \label{extremesignal}
\end{equation}
(Note that at $\xi = 0$, also $\tau = t$). The function $S(\tau)$ is periodic with
the modulation period, and is strictly positive for $3/2 < \hat{\nu}^{2} < 2$,
while for $\hat{\nu}^{2} < 3/2$ it is partly positive and partly negative,
vanishing at the points $\tau = \zeta$ for which
\begin{equation}
\cos(\nu\zeta) = \frac{1 - \hat{\nu}^{2}}{\sqrt{1 - \hat{\nu}^{2}/2}}.
\end{equation}

In the following, we will call the possibly non-sign definite function $S$ the `envelope', to distinguish it from (half) the `amplitude' $|S|$ which is nonnegative.
In fact, in polar coordinates, the signal can be written like
\[
s_{0}(t) = 2|S(t)| \cos \left(\omega_{0}t + \pi \cdot \left(1 - \text{sign} \, S \right)/2 \right).
\]

\begin{figure}[h]
\begin{center}
\includegraphics[width = 0.4\textwidth]{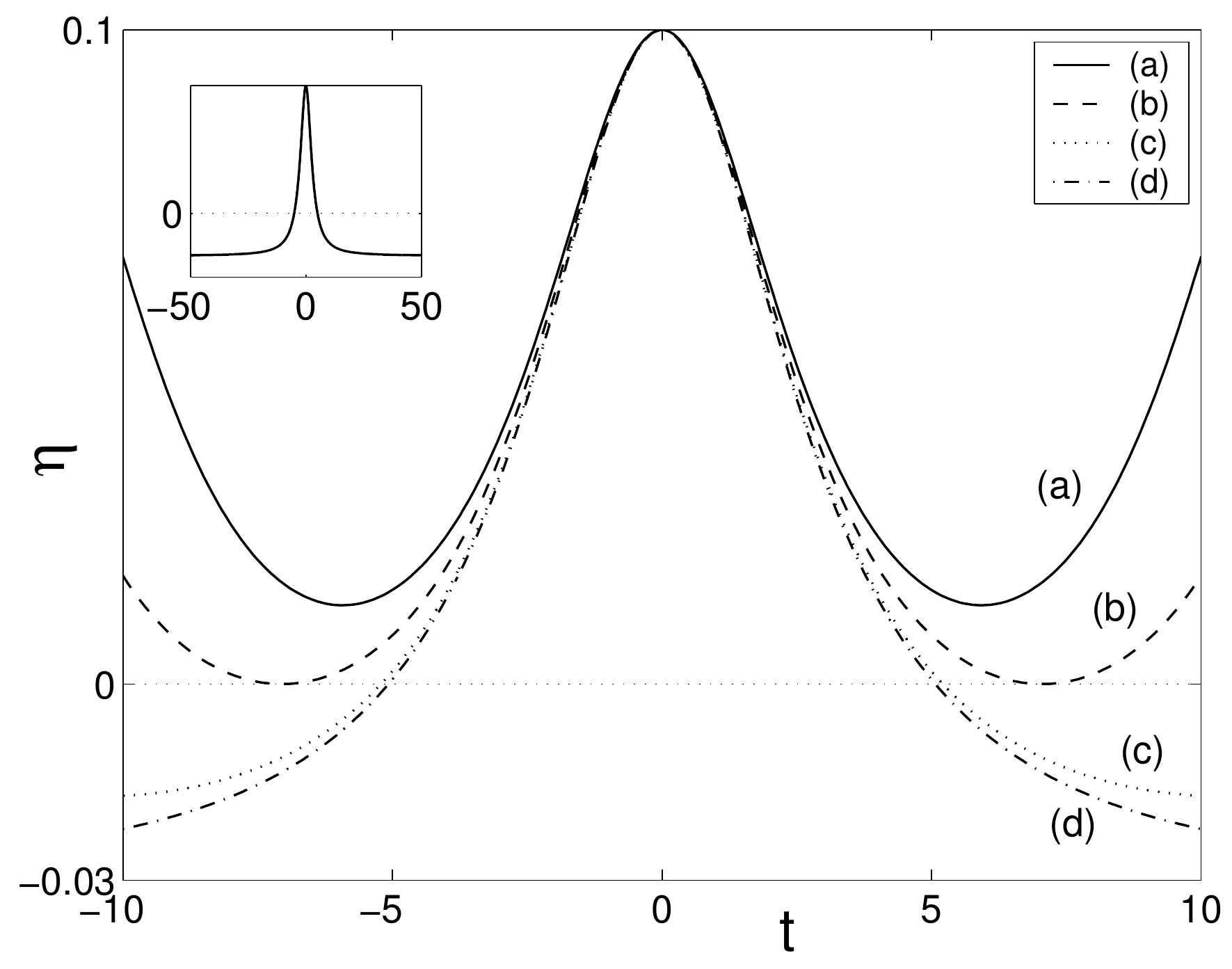}
\end{center}
\vspace*{-0.5cm}
\caption{Plot of (twice) the extreme envelope, the function $2S$ (the factor of $2$ for ease of relating the plots to those of the physical waves: $\eta = 2S \cos \left(\omega_{0}t \right)$) for three different values of $\hat{\nu}$: (a) $\hat{\nu} = 1.3$, (b) $\hat{\nu} = \sqrt{3/2}$, (c) $\hat{\nu}=1$. In case (a) there are
no phase jumps, while in case (c) there are, namely the zero's of $S$, and (b) is the separating case. Also shown (repeated in the inset on a larger horizontal scale) is the graph (d) of the limiting function $2S$ for $\hat{\nu} \rightarrow 0$ which is a soliton on negative finite background.} \label{5}
\end{figure}

This makes it explicit that, since $2|S|$ is the amplitude, the signal has a phase jump $\pi$ at the points where $S$ vanishes. As stated above, this happens for parameter values $\hat{\nu}^{2} < 3/2$, which is for part of the B-F instability interval. This is illustrated in Fig.~\ref{5} which shows plots of $S(\tau)$ for three different values of $\hat{\nu}$ that all lie inside B-F instability interval $(0,\sqrt{2})$, namely $\hat{\nu} = 1$, $\sqrt{3/2}$, $1.3$. Observe the limiting case $\hat{\nu} = \sqrt{3/2}$ which is the value for the signal that separates the strictly positive functions $S$ from the indefinite functions, and the limiting case $\hat{\nu} \rightarrow 0$ for which the function $S$ becomes like a soliton on finite (negative) background. In the next section, we show that it is, in fact, a real soliton; the negative value of the finite background is precisely the value of the wave field at infinity, $2r_{0}$.

In the next section, we will further analyse the extreme signal. For here it is sufficient to use the appearance of a phase jump in the extreme signal as a
motivation to investigate what the related phenomena are for the total wavefield. This leads to the finding of wave dislocations, as we will detail further on.

\subsection{Wave dislocation}   \vspace*{0.25cm}

The phenomenon of wave dislocation is known to appear in surface waves \cite{Tan95}, but little attention seems to have been given to it. It is shown to illustrate the performance of numerical codes in \cite{Hen99,Dold92}, and was explained as a result of vertical vorticity in \cite{Cos99}. In optics literature, see e.g.
\cite{BalEA00,Bas95,Fre99,Gbur02,NyeBer74,PopDog02}, the phenomenon of wave dislocation is often called phase-singularity, and has been measured experimentally and
simulated as reported in \cite{BalEA00}; in the linear case, the intensity amplification is of course lacking.

\begin{figure}[h]
\begin{center}
\includegraphics[width = 0.45\textwidth]{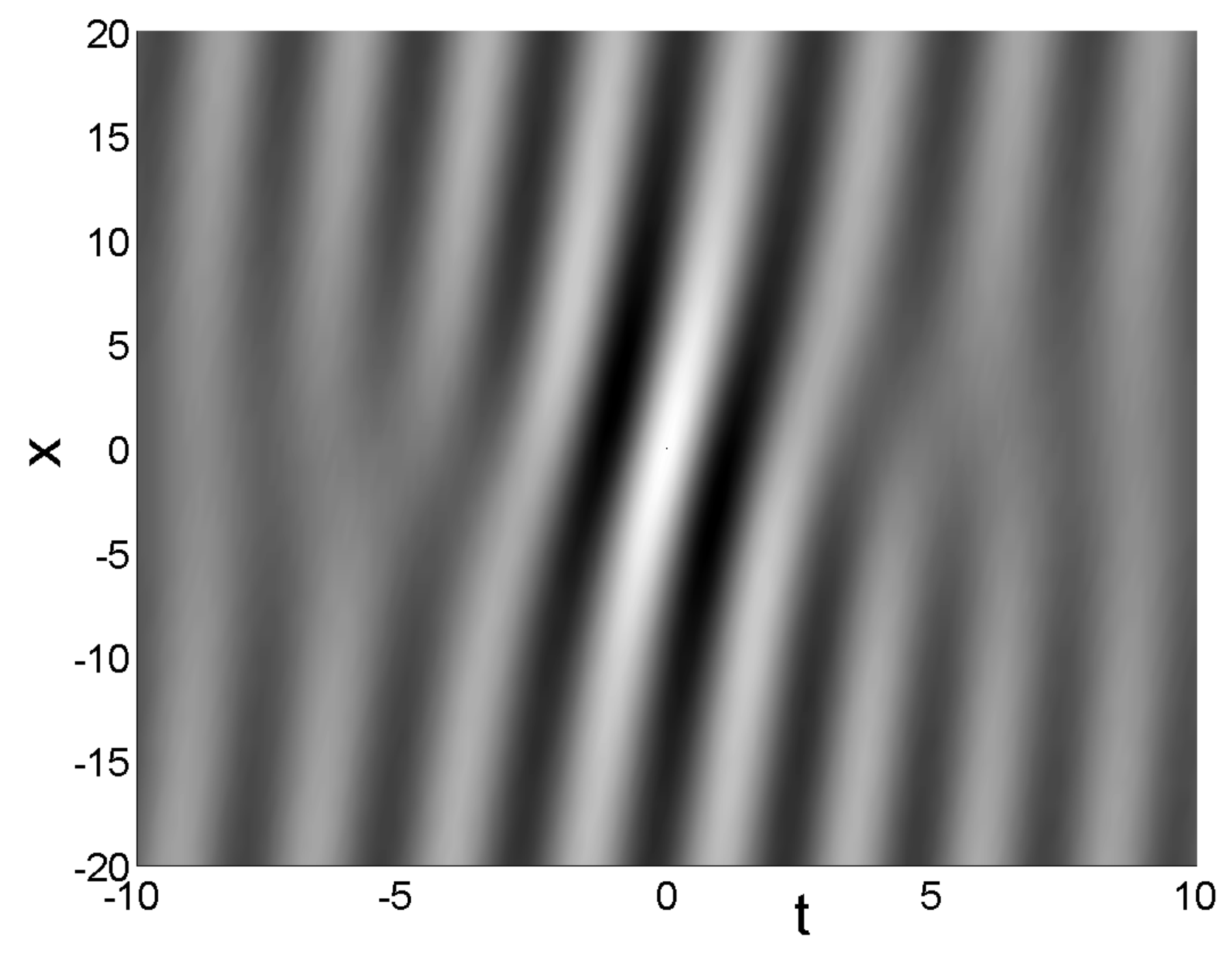}
\end{center}
\vspace*{-0.5cm}
\caption{Density plot of the wavefield showing in more detail the wave splitting at the left of the origin and wave merging at the right. Horizontally is the physical time $t$ shifted with the group velocity, and vertically the physical space $x$.} \label{6}
\end{figure}

A wave dislocation, a phenomenon of merging of two waves into one, or the inverse process of splitting of one wave into two, is a consequence of the phase singularities in SFB. Fig.~\ref{6} is an amplification of part of the wavefield shown in Fig.~\ref{3} (left) to show the wave splitting and merging in more detail.

\begin{figure}[h]
\vspace*{-1cm}
\begin{center}
\includegraphics[width = 0.5\textwidth]{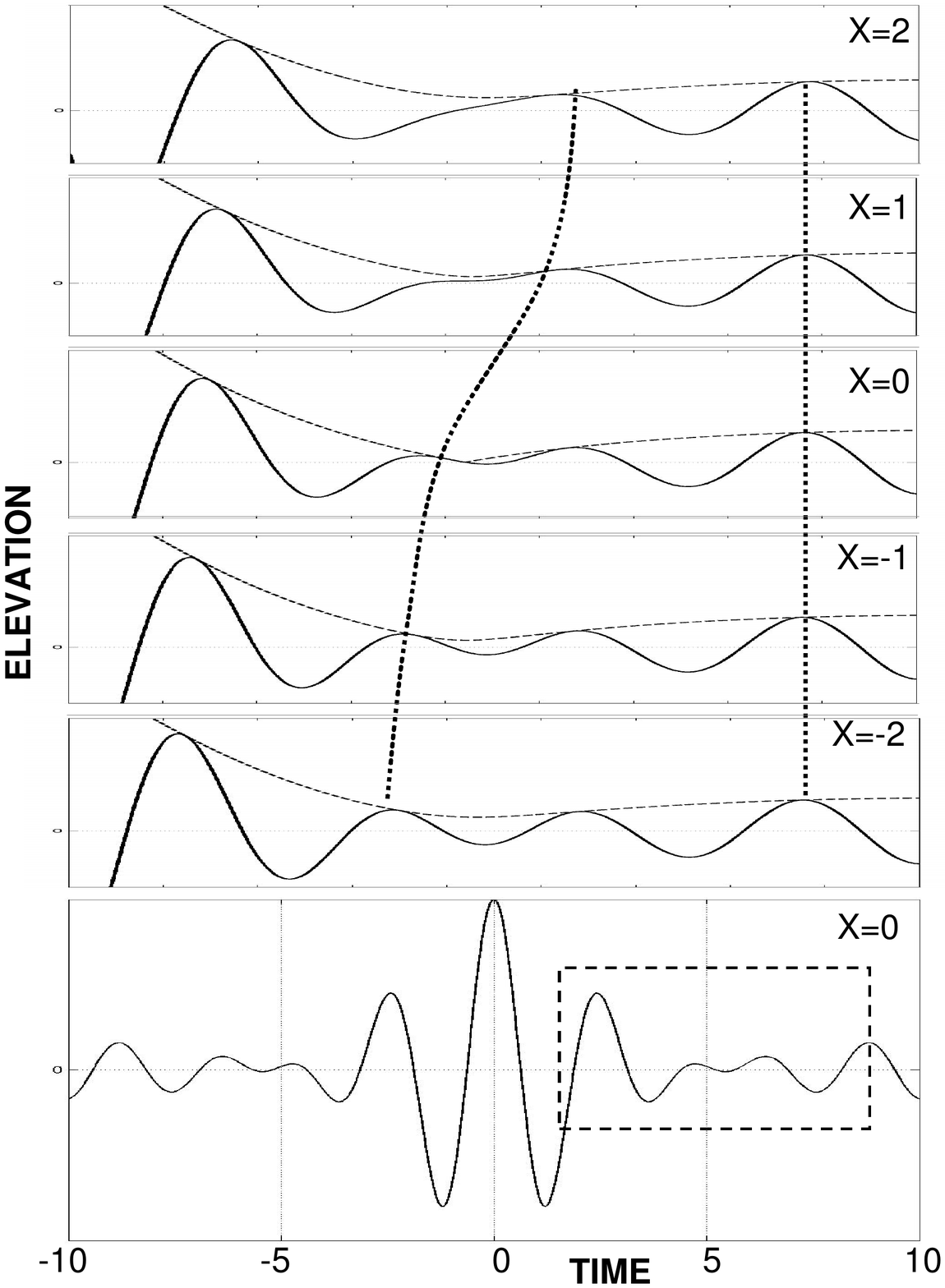}
\end{center}
\vspace*{-1cm} 
\caption{(Above) Slices at various positions near the wave dislocation at $(x,t) = (0,\zeta)$ (with $\zeta > 0$) were wave merging takes place of the solution depicted in Fig.~\ref{6}. Shown are plots at different positions $x$, increasing from negative values below to positive values upwards, of the wave elevation (solid lines) and amplitude (dashed lines) as a function of (shifted) time. The vanishing of the envelope at $(0, \zeta)$ leads to the flattening of the surface in the focusing region, resulting in the disappearance of one wave when crossing $x = 0$ to positive time, as shown by the curve connecting two wave crests. When the sign of the vertical axis is reversed, wave splitting is seen that happens at the point $(0, -\zeta)$. (Below) The manifestation of the phase singularity shown in the time signal at the extreme position; the box indicates the waves in the signal which are visible in the middle of the picture above.} \label{7}
\end{figure}

In Fig.~\ref{7} this wave splitting and merging is shown to be related to the vanishing of the amplitude of the extreme wave signal $S(\tau)$ at $t=\zeta$ for values $\hat{\nu} < \sqrt{3/2}$. In Fig.~\ref{7} are shown plots of the time signal of $\eta_\text{SFB}$ and its envelope (dashed) at various positions near the extreme
position $x = 0$. At $x = 0$ the amplitude vanishes at $t = \zeta$, described as `disapearance of waves' in \cite{Tan95}; the relation to the singular behaviour will be described further on.

\subsection{Dispersion plots} \vspace*{0.25cm}

We will now describe the phenomenon of wave dislocation in a more analytical way by showing that it is related to large changes in the local wavelength and frequency. This is best illustrated using the phase-amplitude description for the wave field (\ref{physicalSFB}) and introducing local wavenumber and local frequency as usually:
\begin{equation*}
\eta(x,t) \approx 2 a(x,t)\cos \Phi \left(x, t \right), \qquad k(x,t) = \partial_{x} \Phi(x,t); \qquad \omega(x,t) = -\partial_{t}\Phi(x,t).
\end{equation*}

\begin{figure}[h]
\begin{center}
\includegraphics[width = 0.4 \textwidth]{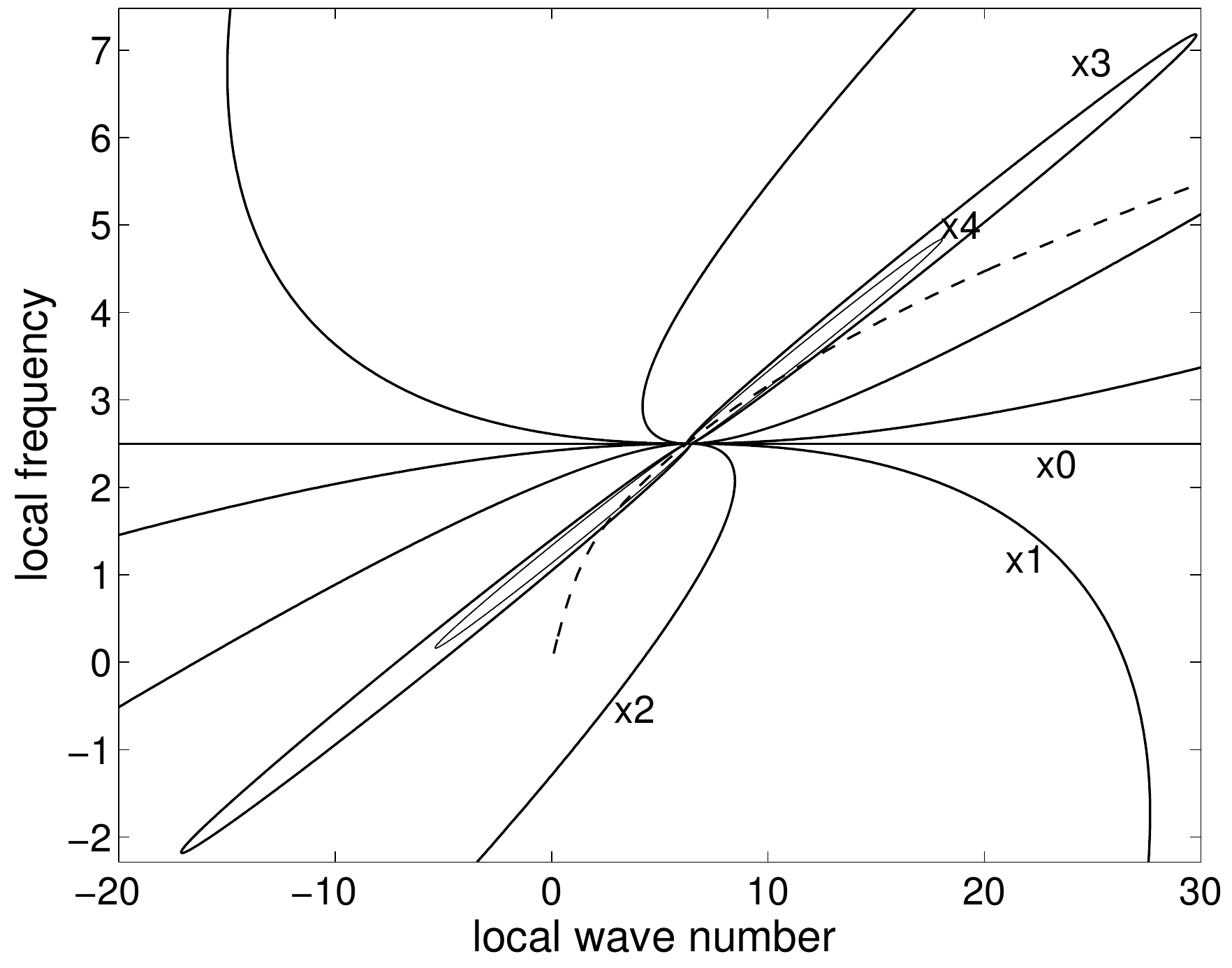} \qquad \qquad 
\includegraphics[width = 0.405\textwidth]{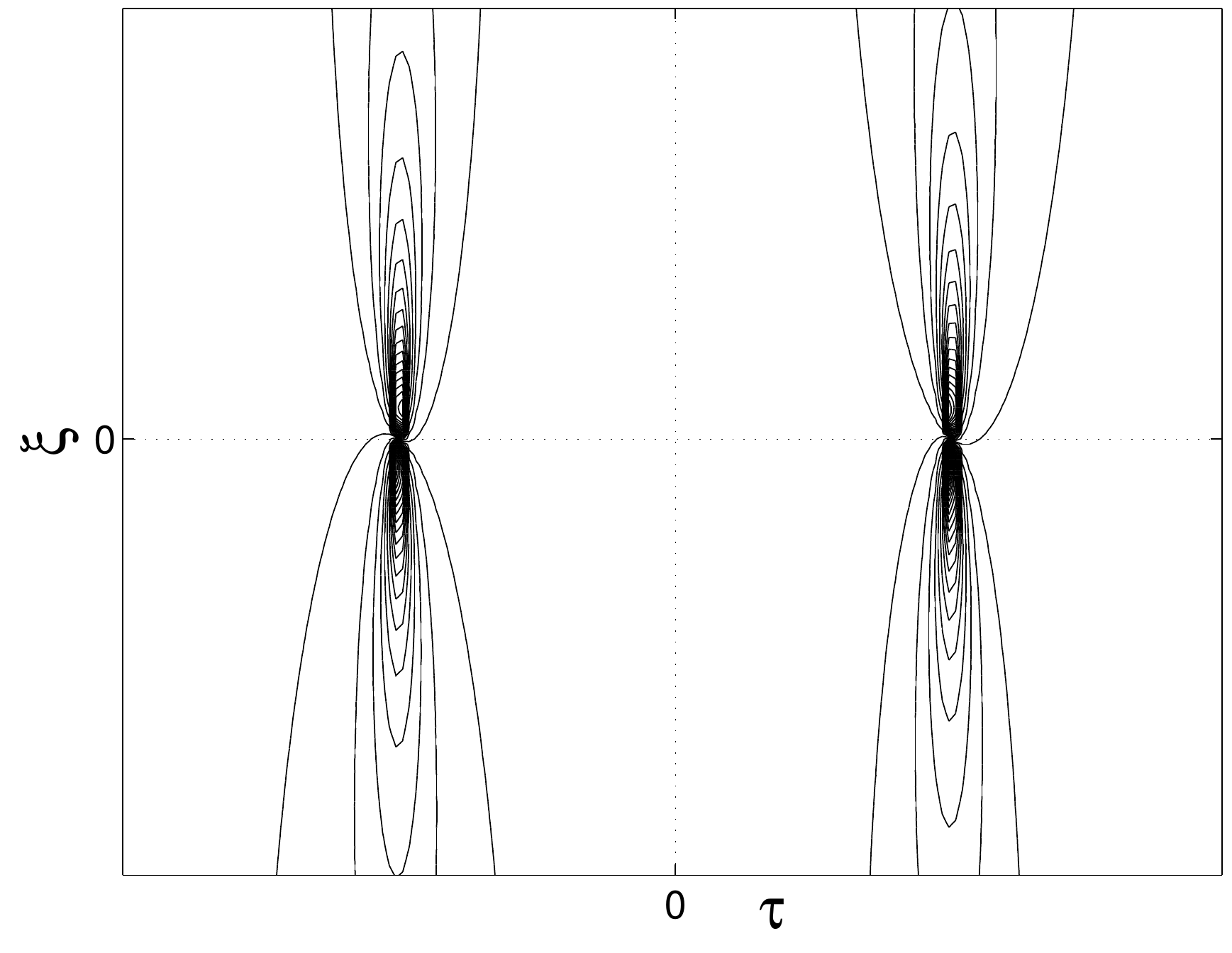}
\end{center}
\vspace*{-0.5cm} 
\caption{(Left) In the dispersion plane of local frequency (vertical) vs local wavenumber (horizontal), trajectories parameterized by $t$ are drawn of $\eta_\text{SFB}$ for various positions, $x = 0 < x_1 < x_2 < \cdots$. The phase singularity is observed for $x = 0$ where the line $\omega = 2.5$ connects $\infty$ and $-\infty.$ The dashed line is the linear dispersion relation $\omega = \Omega(k)$. (Right) Contour plots of the wave number as function of $\tau = t - x/V_{0}$ (horizontal) and $\xi = x$ (vertical). Observe the abrupt change of the wave number for $\xi = 0$ at the wave dislocations: splitting at the left ($t = -\zeta$) and merging ($t = \zeta$) at the right.} \label{8}
\end{figure}

With these definitions, the evolution of the local wavenumber and local frequency can first be investigated in the \textit{dispersion plane} of frequency versus wavenumber. In figure Fig.~\ref{8} (left), for various positions $x$ the trajectories $t \rightarrow (k(x,t), \omega(x,t))$ are shown parameterized by $t$. Observe that for the value $x = 0$ a straight line appears connecting at $\omega=2.5$ the $k$-values $\infty$ and $-\infty$, which describes the phase singularity at the
position and time at which the wave number $k$ becomes unbounded. This phenomena can further be seen from Fig.~\ref{8} (right) where contour plots are drawn for the wave number $k(x,t)$ near the location $\xi = x = 0$. Large and abrupt changes of the local wave number are visible near the wave dislocations at $t=\zeta>0$ (wave
merging) and at $t=-\zeta$ (wave splitting), showing the singular behaviour at the point of the dislocation and justifying the use of the word `phase-singularity'. For later reference we note that the contour lines show that $\partial_{x}k < 0$ for merging, and $\partial_{x}k > 0$ for splitting.

\subsection{Energy exchange and wave discrimination} \vspace*{0.25cm} \label{eewd}

We will now present an explanation of the large amplitude amplification between phase singularities, or better to say, to the separation process of successive intermittent and extreme waves. Therefore we return to the governing NLS equation and rewrite this equation in physical coordinates in the real amplitude $a$ and local wave number and local frequency. The governing equations, obtained by looking at the real and imaginary parts, are the so-called \textit{phase-amplitude equations}. The phase equation is given by
\[
\omega - \Omega(k) = V_0 \beta \frac{\partial_{t}^{2}a}{a} + V_0 \gamma a^{2},
\]
where $\Omega(k)$ is the linear dispersion relation, and the amplitude equation is effectively the energy equation
\[
\partial_{t} E + \partial_{x} \left[V \left(k \right) E \right] = 0, \qquad E = a^{2}.
\]
The phase equation contains the so-called Fornberg-Whitham term $\beta \frac{\partial_{t}^{2} a}{a}$ which becomes singular when the amplitude vanishes; this leads to the singular behaviour in the dispersion plane as described above.

The resulting large amplitudes in between wave-splitting and wave-merging are caused by nonlinear effects, although the wave dislocation itself is actually a linear phenomenon. The amplitude amplification is caused by the interplay between the envelope and the changes in phases of the carrier wave, and can be seen as follows. The energy equation describes conservation of integrated energy density between group lines (see \cite{Whitham}). Stated differently, this equation shows that the change of energy transported with the local group velocity is forced by local changes in wavelength:
\[
\left[\partial_{t} + V \left(k \right) \partial_{x}\right] E + E \partial_{x} V \left(k \right) = 0.
\]

These changes are largest near the dislocations. Since $\partial_{x} V \left( k \right) = \Omega^{\prime\prime} \left( k \right) \partial_{x}k$, and $\Omega^{\prime\prime} \left( k \right) < 0$ for surface wave dispersion, there is a decrease or increase of energy near the dislocations depending on the
character of the dislocation. At wave splitting, it holds $\partial_{x} k > 0$ as shown above, and hence the energy increases while travelling with the
group velocity. The group velocity is for surface wave dispersion less than the phase velocity, and hence there is a nett energy flow crossing the wave at the
merging position from left to right. Reversely, at a position of wave merging $\partial_{x} k < 0$ and there is a nett energy flow from right to left. These two
effects enhance each other and lead to energy increase between a splitting singularity and a merging singularity. This nett transport of energy causes that the waves are divided into intermittent waves and extreme waves. This energy flow leads to changes of the amplitudes of the waves that can become as large
as illustrated in the plots (see also Section~\ref{dsse}). Hence, the process of wave discrimination is caused by energy exchange through successive wave dislocations.

\section{Extreme signals} \label{es}

We now return for a further study to the extreme signal, in particular to the (possibly non sign-definite) envelope function $S$ in (\ref{extremesignal}) as plotted in Fig.~\ref{5}. Of special interest are the cases for which $S$ is partly negative: the cases of phase singularities and wave discrimination, i.e. values 
$\hat{\nu} < \sqrt{3/2}$.

\subsection{Phase plane representation} \vspace*{0.25cm}

\begin{figure}[h]
\begin{center}
\includegraphics[width = 0.4\textwidth]{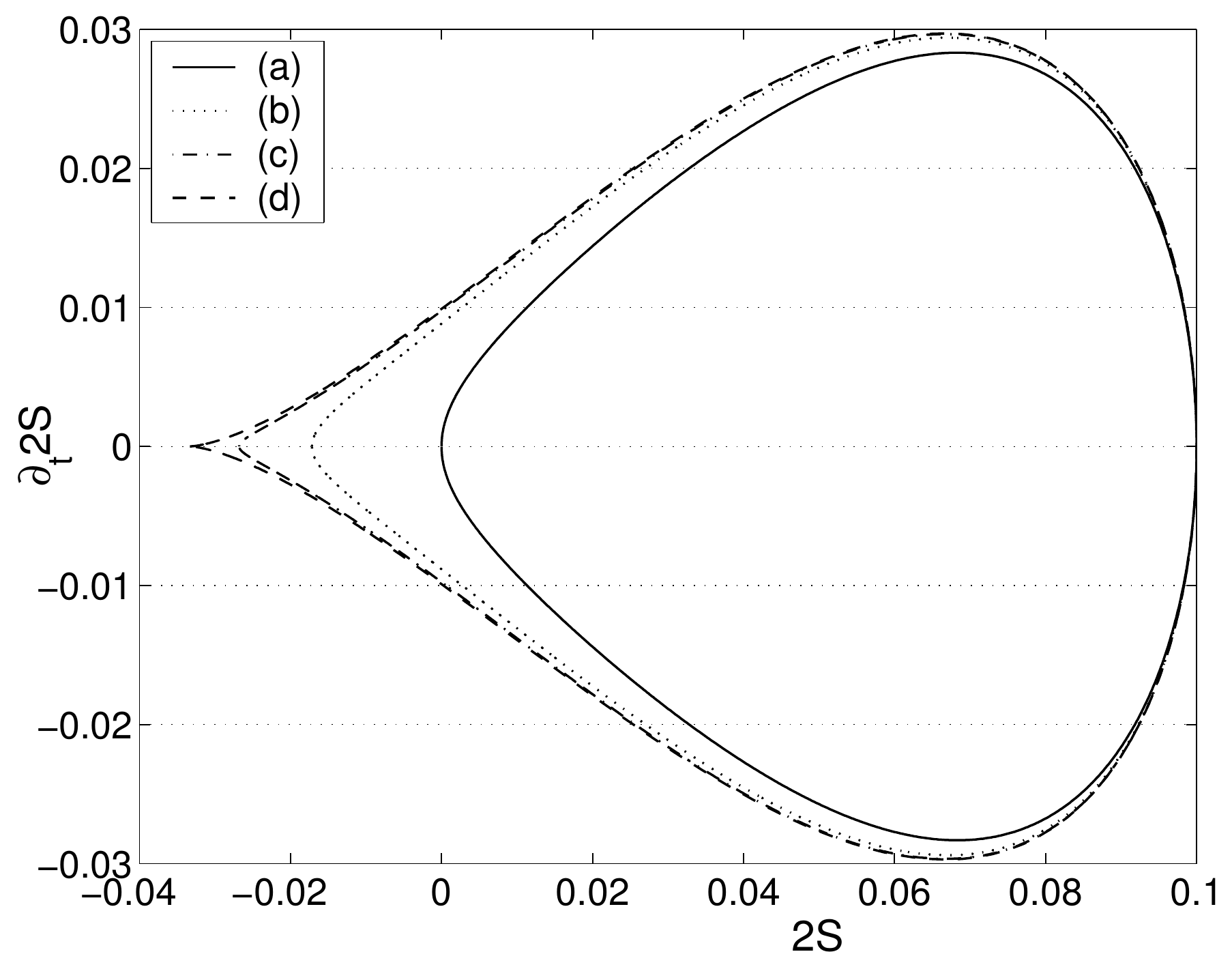} \qquad \qquad 
\includegraphics[width = 0.4\textwidth]{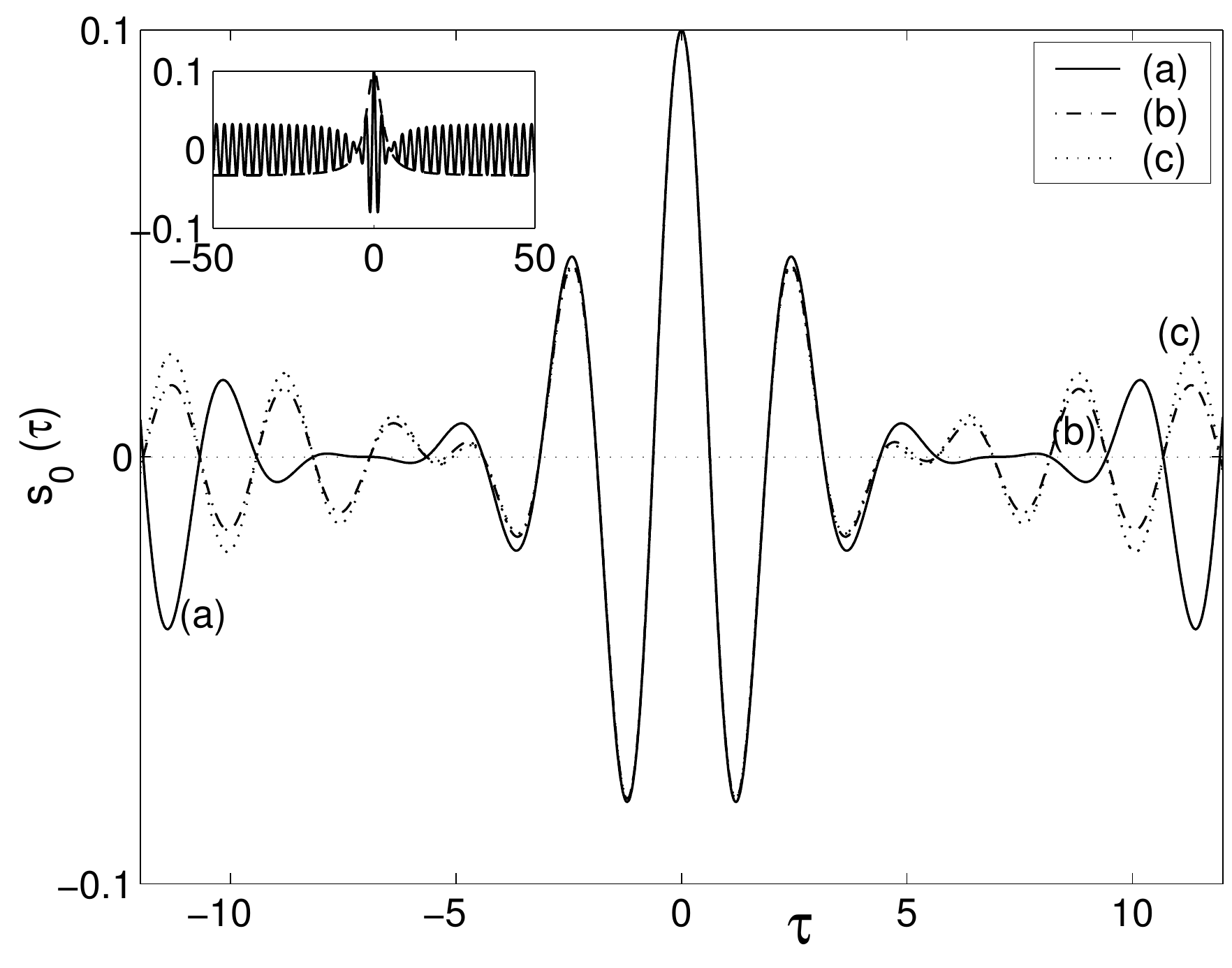}
\end{center}
\vspace*{-0.5cm}
\caption{(Left) In the phase plane of $2S$ (horizontally) vs $\partial_{t} \,2S $ (vertically) the trajectories are plotted of $S$ as function of time for three values of the normalized modulation frequencies $\hat{\nu}$: $\sqrt{1/2}$, $1$, $\sqrt{3/2}$ (from outside to inside). The `limiting' case 
$\hat{\nu} \rightarrow 0$ is approximated by $2S$ for very small value of $\hat{\nu}$ and leads to the most outward curve. (Right) Plot of the signals $s_{0}(\tau) = 2S(\tau) \cos\left( \omega_{0} \tau \right)$ for the same values of $\hat{\nu}$ as in the left picture. The resemblance of the extreme waves for all these different values shows once more the generic form of these extreme profiles. In the inset the limiting case $\hat{\nu} \rightarrow 0$ is shown on a much larger time interval.} \label{9}
\end{figure}

We will start to draw the `phase portrait' of $S$, that is to say, parameterized by $t$, the function values are plotted against the values of the derivative in figure Fig.~\ref{9} (left), leading to a trajectory in the phaseplane of $\partial_{\tau} \, 2S$ versus $2S$. The different trajectories correspond to different values of the normalized modulation frequencies $\hat{\nu} = \sqrt{1/2}$, $1$, $\sqrt{3/2}$. The corresponding signals $s_{0}(\tau)$ for those values are plotted in Fig.~\ref{9} (right).

\begin{figure}[h]
\begin{center}
\includegraphics[width = 0.6\textwidth]{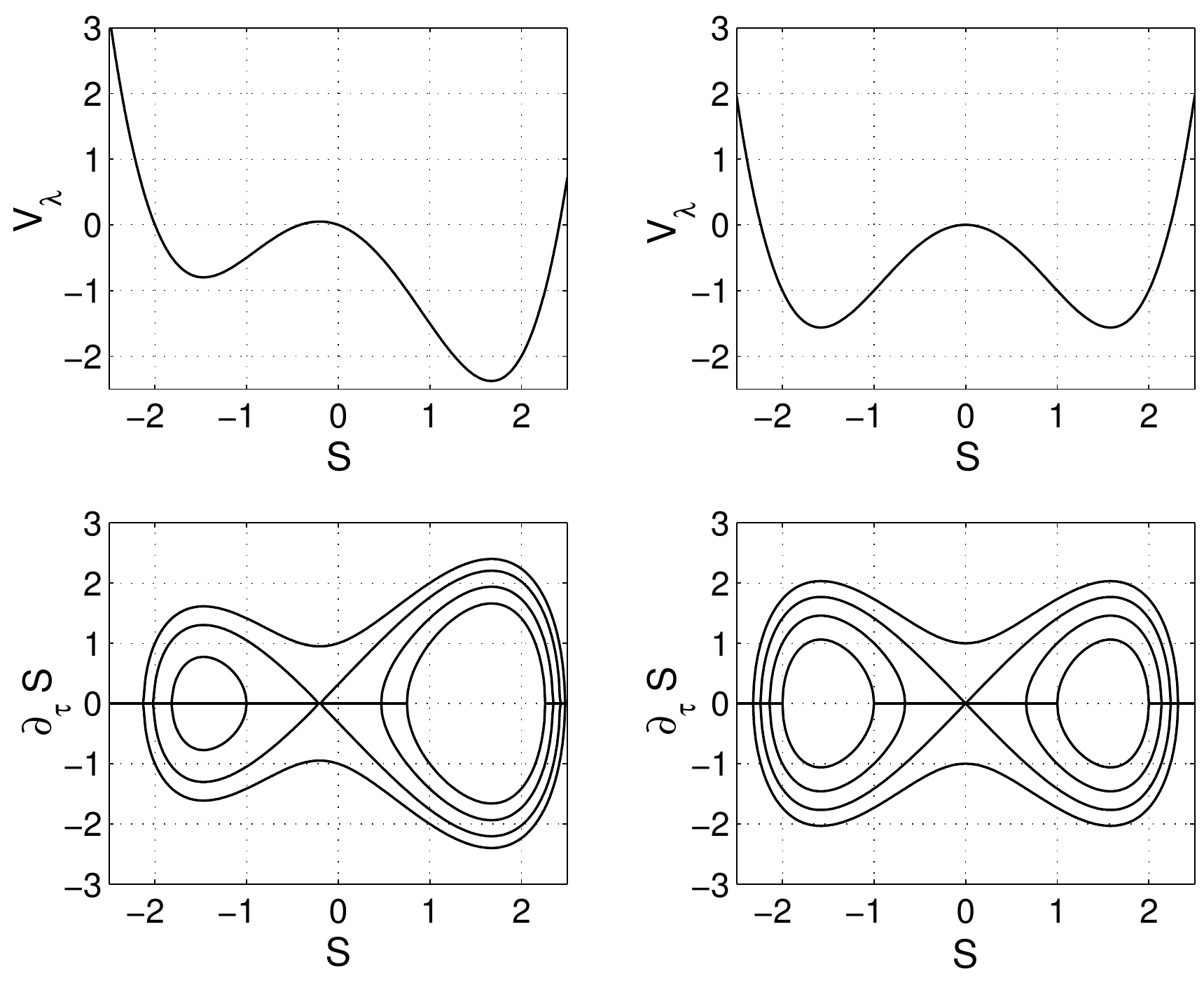}
\end{center}
\vspace*{-0.5cm}
\caption{(Left column) For normalized variables $\gamma = \beta = 1$, and $\kappa = 2.5$ and $\lambda = 0.5$, the phase plane is shown with the potential $V_{\lambda}$ above it. The interesting extreme signals, of amplitude $M/2$, are obtained for $\mathcal{E} = V_{\lambda} \left(M/2 \right) > 0 $, and correspond to the periodic motions that are mainly in the positive half plane but have some negative values. Observe the homoclinic orbit through the equilibrium (the local maximum) which is $-2r_0$. These trajectories correspond to those shown in Fig.~\ref{9} (left). (Right column) For the same variables, the symmetric potential $V_{0}$ with $\lambda = 0$ is shown, with below it the phase plane. The steady-state profiles of NLS, nonlinear normal mode, periodic amplitude oscillations and the soliton correspond to the (positive) equilibrium, the periodic oscillations around it, and the homoclinic orbit through $(0,0)$.} \label{10}
\end{figure}

Plots as these are often found in systems of one degree of freedom in classical mechanics. Indeed, it can be verified that the extreme signal
satisfies Newton's equation
\begin{equation}
\beta \partial_{\tau}^{2} S + \gamma S^{3} - \kappa S - \lambda = 0,  \label{Newtonsignal}
\end{equation}
with parameters $\lambda, \kappa$ given by
\begin{equation}
\lambda = \gamma r_{0}^{3} (2 - \hat{\nu}^{2} ), \qquad \kappa = \gamma r_{0}^{2} (3 - \hat{\nu}^{2}).  \label{lambdakappa}
\end{equation}
When written like $\beta \partial_{\tau}^{2} S + \partial V/\partial S = 0$, the equation can be interpreted as the equation of motion of a particle with mass
$\beta$ subject to a conservative force with potential $V_{\lambda}(S) = \frac{\gamma}{4} S^{4} - \frac{1}{2} \kappa S^{2} - \lambda S$ (where we made, for later reference, the dependence of the potential on $\lambda$ explicit in the notation). The trajectories in the phase plane are therefore level lines with 
$\mathcal{E} > 0$ of the total (conserved) `energy'-like quantity:
\[
\mathcal{E} = \frac{\beta}{2} \left(\partial_{t} S\right)^{2} + V_{\lambda}(S), \qquad \partial_{\tau} \mathcal{E} = 0.
\]
The potential and the corresponding phase plane are shown in Fig.~\ref{10} (left column).

The extreme signal can be written down in the form of an elliptic integral%
\[
t = \sqrt{\beta} \int^{S} \frac{ds}{\sqrt{2(\mathcal{E} - V_{\lambda}(s))}}%
\]
where the total energy can be expressed in the maximal amplitude $M$ as $\mathcal{E} = V_{\lambda} \left(M/2 \right)$.

\subsection{Signal potential as unfolded NLS-steady state potential} \vspace*{0.25cm}

It is illustrative to compare the above phase plane analysis with `standard' soliton solution of NLS. This solution and periodic modulations of the nonlinear monochromatic mode are found from the NLS equation by looking for solutions of the form in separated variables 
\[
A = e^{-i \kappa\xi} f\left(\tau \right)
\]
with $f$ a real function and $\kappa$ a real number to be determined together with $f$. In physical variables, the corresponding wave field looks like 
\[
\eta_{ss}(x,t) = 2 f(t - x/V_{0}) \cos \left((k_{0} - \kappa)x - \omega_{0} t \right),
\]
i.e. a monochromatic carrier wave with adapted dispersion, modulated with a profile function $f$ that steadily progresses with the group velocity. For this reason, solutions of this kind are usually called `steady state' solutions. Substitution of the above Ansatz into the NLS equation leads to the ODE for $f$%
\[
\beta \partial_{\tau}^{2} f + \gamma f^{3} - \kappa f = 0.
\]

This is recognised as Newton's equation with a potential $V_{0}(f) = \frac{\gamma}{4}f^{4} - \frac{\kappa}{2}f^{2}$. The potential and the corresponding phase plane trajectories, satisfying the `energy' conservation $\beta\left(\partial_{\tau} f \right)^{2} + V_{0} \left(f \right)=$ constant, are shown in Fig.~\ref{10} (right column) for the (only) interesting case of $\kappa>0$. In the phase plane, the nonlinear monochromatic mode, for which the potential is minimal, and periodic
oscillations around this monochromatic mode, and the soliton solution as the homoclinic orbit through $f=0$, are to be noted. Explicit formul{\ae} in terms of elliptic integrals can be written down as before.

Compared to the potentials $V_{\lambda}$ defined above that describe the extreme signals for $\lambda > 0$, the potential $V_{0}$ considered here for the standard NLS is precisely the potential $V_{\lambda}$ for the special case $\lambda = 0$. Hence $V_{\lambda}$ can be seen as an unfolding of the potential $V_{0}$. For $\lambda > 0$ the symmetry that is present in $V_{0}$ is broken, as is seen in the plot of the potential and in the phase plane. Although, from this point of view, the unfolding is a rather simple phenomenon, on the level of solutions of the NLS equation, this unfolding has large consequences for the governing dynamics in $\xi$, which shows itself in the very different NLS-solutions: when $\kappa$ and $\lambda$ are related to the quantities $r_{0}$ and $\hat{\nu}$ according to (\ref{lambdakappa}), the corresponding SFB-solutions are obtained.

\subsection{Phase singularity related to extreme signal} \vspace*{0.25cm}

From the equation (\ref{Newtonsignal}) satisfied by $S$, we can extract more information about the solution in a neighbourhood of the extreme position $\xi = 0$. 
To that end first note that at $\xi = 0$, $A = S(\tau)$. Then since $A$ satisfies the NLS equation the restriction to the point $\xi = 0$ implies that \begin{equation*}
\left. \partial_{\xi} A \right \vert_{\xi = 0} = -i \gamma \left(\frac{\beta}{\gamma} \partial_{t}^{2}S + S^{3}\right) 
= -i \gamma \left(\lambda_{2}S + \lambda_{1}\right)
\end{equation*}
leading to
\[
A = e^{-i \gamma \lambda_{2} \xi} S(\tau) - i \gamma \lambda_{1} \xi \qquad \text{near} \qquad \xi = 0.
\]
Writing $\Phi$ for the phase of $A$, $A = |A| e^{i\Phi}$, it holds that
\[
\Phi = \tan \left(\sigma\xi \right) + \frac{\xi}{\cos \left(\sigma\xi \right)} \frac{\lambda}{S(\tau)},
\]
confirming that $\Phi = 0$ at $\xi = 0$, while a simple calculation leads to
\[
\left. \partial_{\xi} \Phi \right\vert_{\xi = 0} = \frac{\lambda}{S(\tau)}.
\]

This shows once again the appearance of the phase singularity, now explicitly and directly relating the zero's of $S$ to the singularity in the change of phase
of $A$, and hence in the phase of $\eta_\text{SFB}$, while crossing the extreme position at times near the events of wave splitting/merging.

\section{Down-stream spectral evolution} \label{dsse}

In Subsection~\ref{eewd} we have used the energy equation to investigate the flow of energy in the physical time-space near the phase singularities. Another common
way to study the deformations is to investigate the evolution of the spectral components. In the present case of a signalling problem, this means that we Fourier analyse the signal at each position and follow the individual Fourier coefficients for increasing position. Although this information is not easily related to actual changes of the physical signal, it contains the same information.

To find the spectral evolution, we write the solution as a superposition of contributions at the carrier frequency $\omega_{0}$ and at sidebands of all orders: $\omega_{0} \pm m\nu$ for $m = 1,2, \dots$:
\begin{equation*}
\eta_\text{SFB} \left(x, t\right) = \sum_{-\infty}^{\infty} c_{m}(x)\exp\left(i \left(\omega_{0} + m \nu \right) t \right) + \text{cc}
= \sum_{-\infty}^{\infty} a_{m}(x) \cos \left((\omega_{0} + m\nu )t + \phi_{m}(x) \right)
\end{equation*}
with real coefficients $a_{m}$ and phase factors $\phi_{m}$ or with complex-valued coefficients $c_{m}$. The energy contents of the central frequency and the sidebands are defined by
\begin{align*}
e_{0}(x) &= |c_{ 0}(x)|^{2} =\frac{1}{4}a_{0}^{2}(x), \\
e_{m}(x) &= |c_{-m}(x)|^{2} + |c_{+m}(x)|^{2} = \frac{1}{4}[a_{-m}^{2}(x)+a_{+m}^{2}(x)] \qquad \text{for} \quad m \geq 1.
\end{align*}
For SFB, the asymptotic behaviour is known to consist of only the carrier wave, and hence
\[
\textnormal{for} \quad x \rightarrow \pm \infty: \quad a_{0}(x) \rightarrow 2r_{0}, \qquad 
a_{m} \left(x \right) \rightarrow 0 \quad \textnormal{for} \quad m=1,2,\dots.
\]
The Fourier coefficients for the sidebands are given by
\[
c_{m} \left(x \right) = \frac{1}{T} \int_{-T/2}^{T/2} \eta_\text{SFB}(x,t) \exp\left(-i (\omega_{0} + m \nu) t\right) \, \text{d}t,
\]
where $T=T_{\text{mod}}$. For the spatial evolution, the total `energy', $\int\eta^{2}/2 \, \text{d}t$, is conserved which implies that the sum of the energy
contents of the modes is conserved:%
\[
\frac{\text{d}}{\text{d}x} \sum_{m = 0,1,2,\dots} e_{m}(x) = 0,
\]
and hence equal to the asymptotic value $r_{0}^{2}$.

\begin{figure}[h]
\begin{center}
\includegraphics[width = 0.3\textwidth]{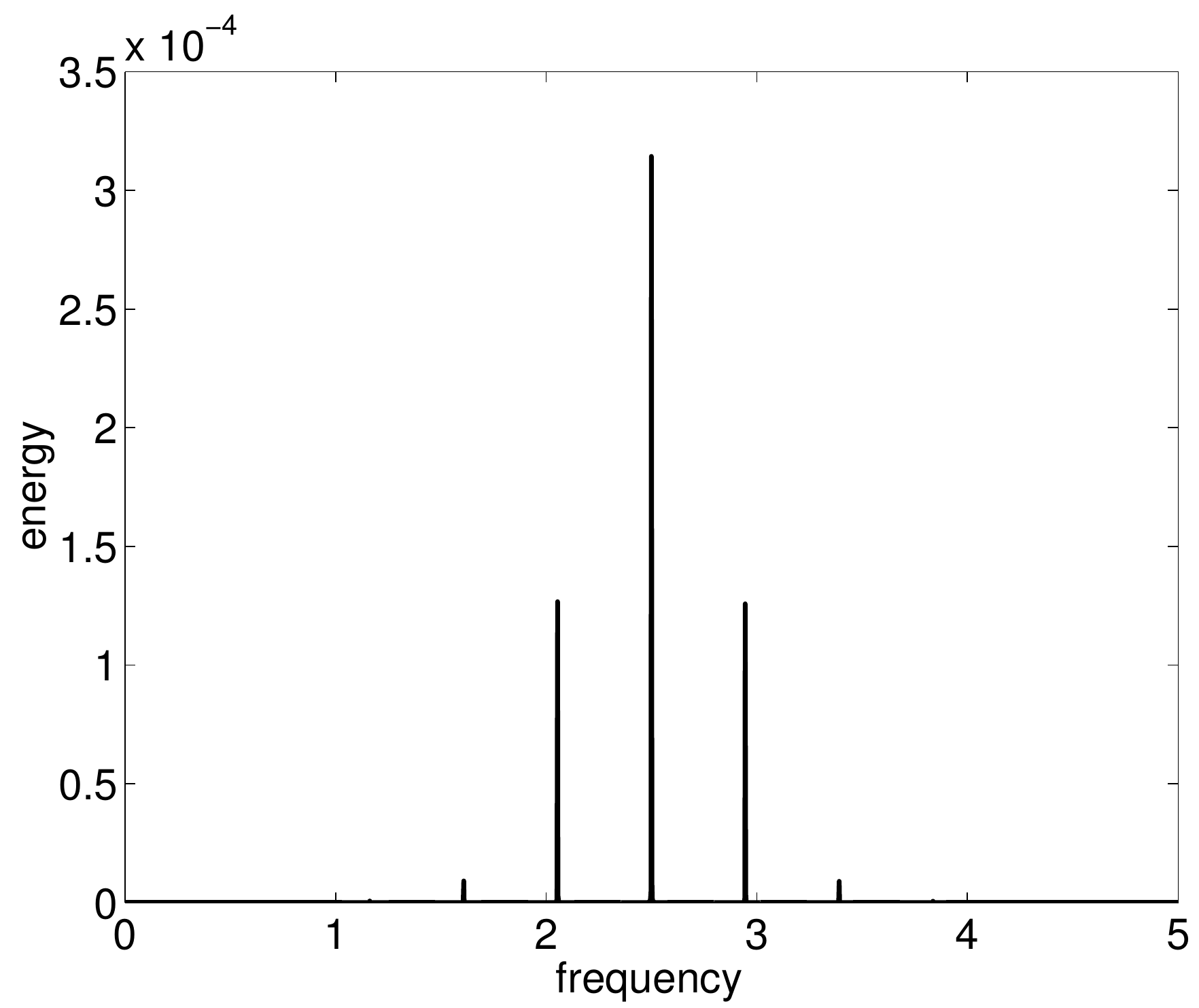} \quad
\includegraphics[width = 0.3\textwidth]{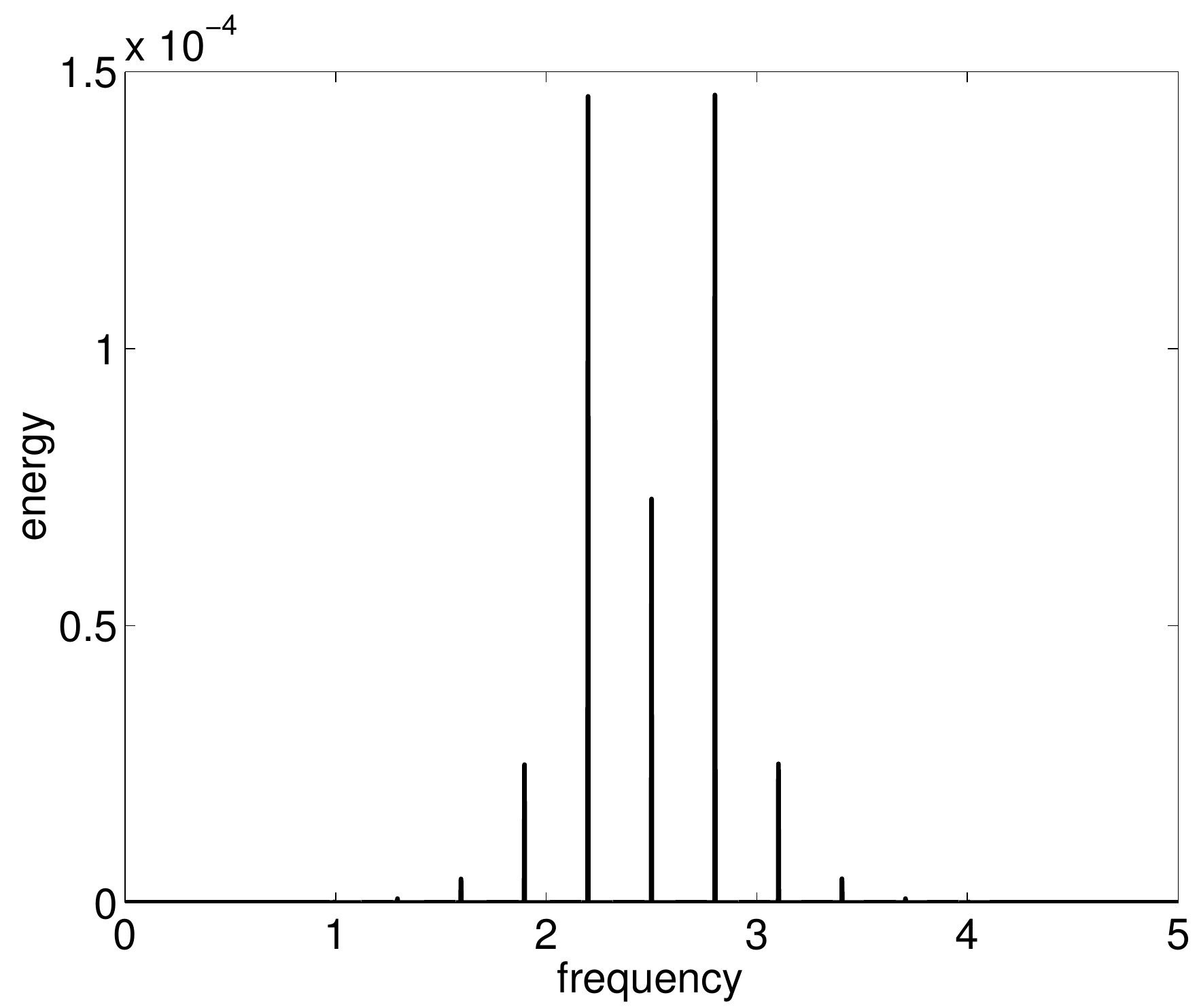} \quad
\includegraphics[width = 0.3\textwidth]{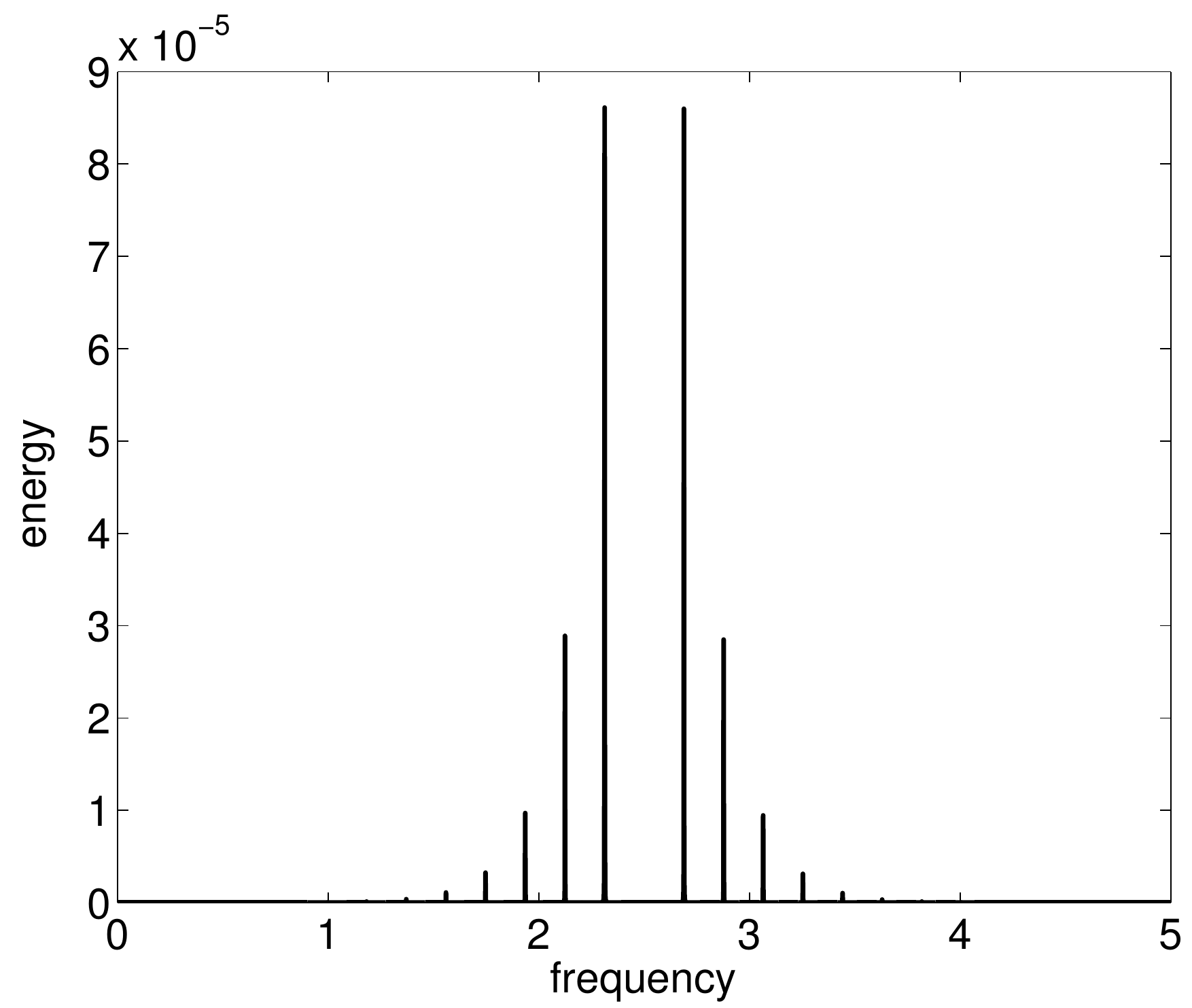} 
\end{center}
\vspace*{-0.5cm} \caption{Fourier spectrum of the extreme signal of $\eta_\text{SFB}$ for $\hat{\nu} = \sqrt{3/2}$, $1$ and $\sqrt{1/2}$ from left to right. 
It is to be noted that the energy content that is asymptotically concentrated in the carrier frequency $\omega_{0}=2.5$ has been redistributed over many sidebands. 
In the special case $\hat{\nu} = \sqrt{1/2}$ the energy at the carrier frequency vanishes precisely.} \label{11}
\end{figure}

\begin{figure}[h]
\begin{center}
\includegraphics[width = 0.3\textwidth]{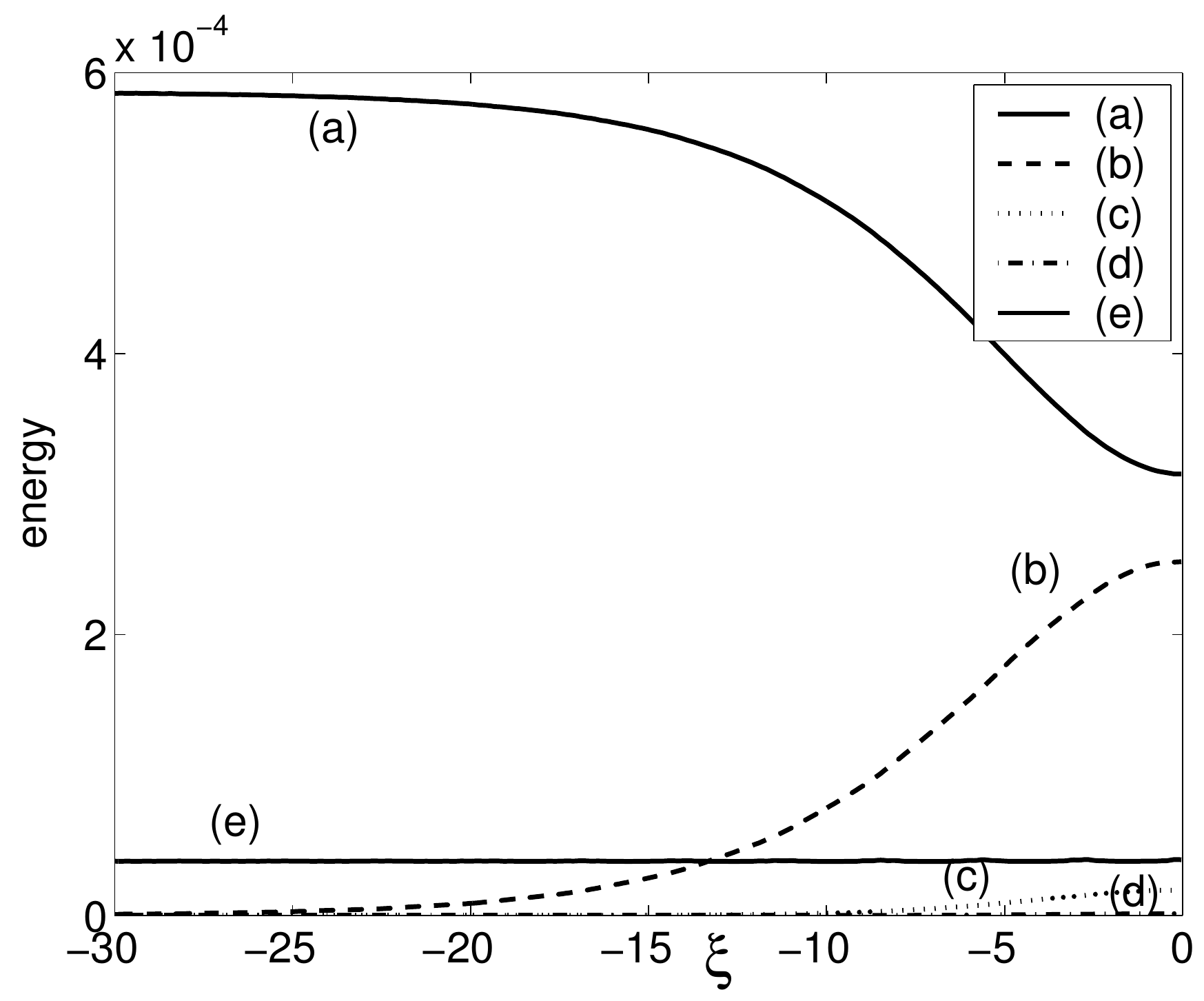} \quad
\includegraphics[width = 0.3\textwidth]{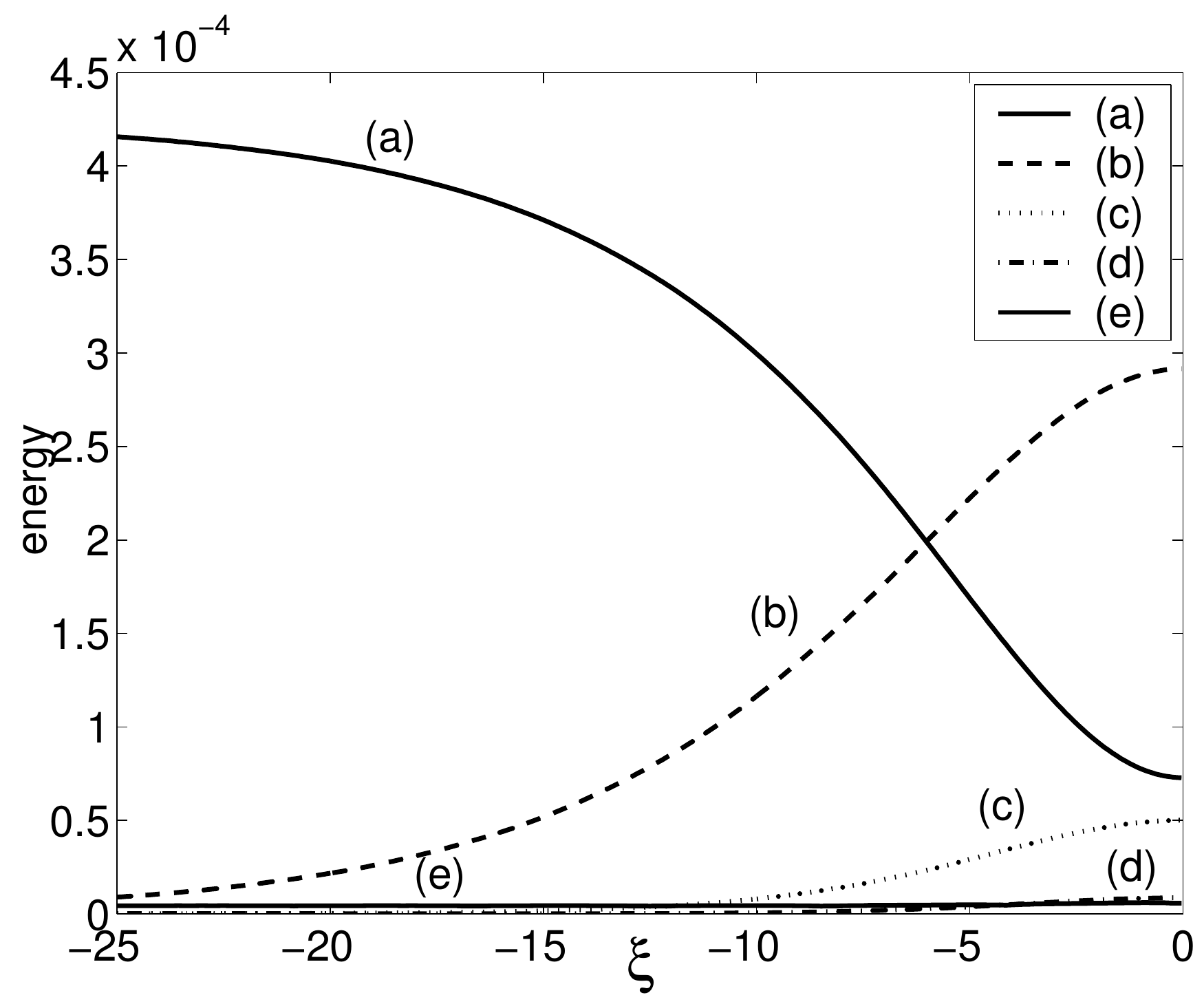} \quad
\includegraphics[width = 0.31\textwidth]{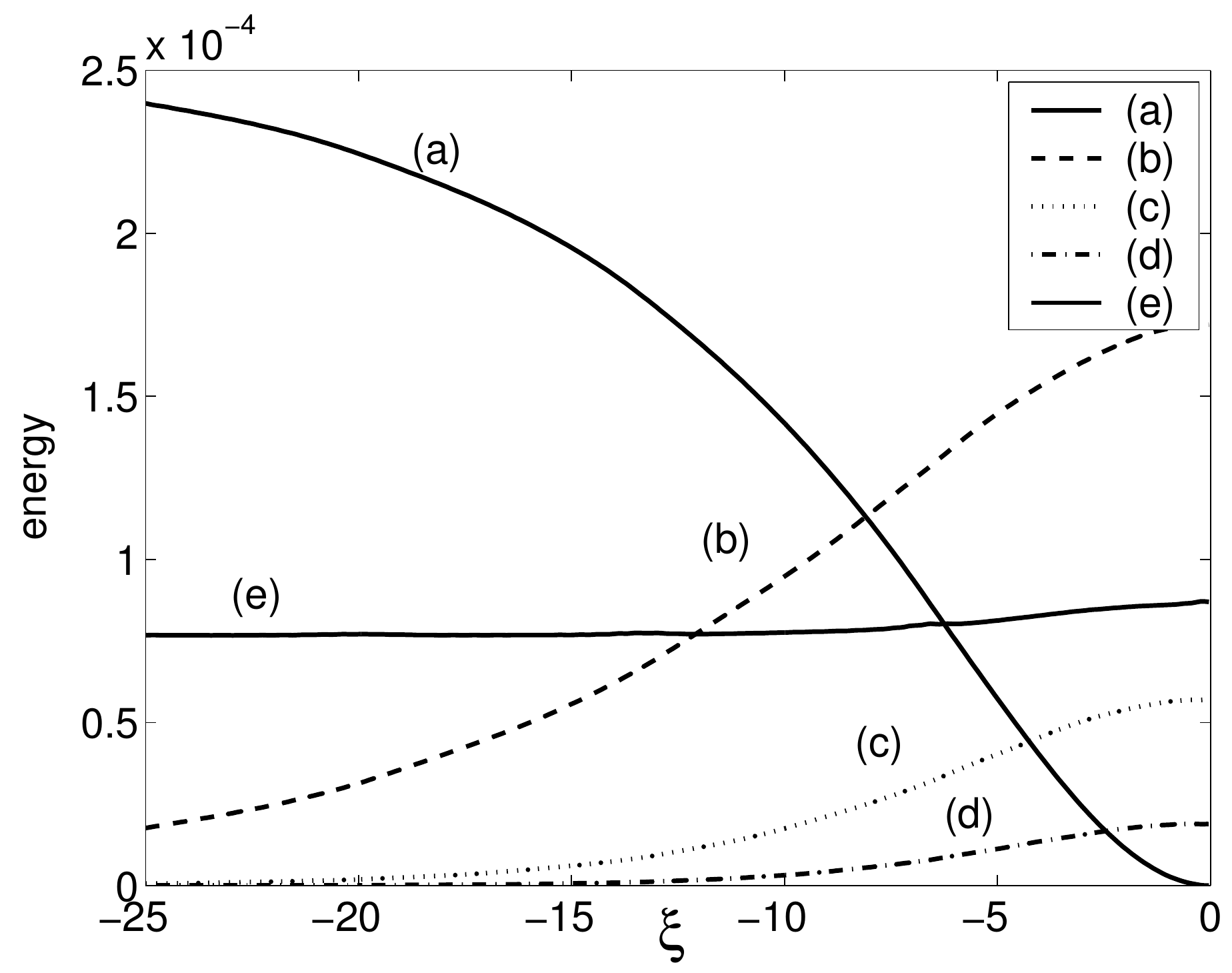} 
\end{center}
\vspace*{-0.5cm} \caption{For $\hat{\nu} = \sqrt{3/2}$, $1$, $\sqrt{1/2}$ from left to right are shown the evolution of the energy of the carrier frequency $e_{0}$ 
(a), of the three lowest sidebands $e_{m}$, $m = 1,2,3$ (b), (c) and (d) respectively, and of the remaining energy denoted and defined by 
$e_{\ast} \left(x\right) = \Sigma_{m>3} e_{m}(x)$, (e).}
\label{11a}
\end{figure}

In Fig.~\ref{11} the Fourier spectrum of the extreme signal is plotted for three cases. The spatial dynamics of the energy in sidebands during the downstream evolution is shown in figure Fig.~\ref{11a}. Plotted are the energy content of the carrier wave and of the first few sidebands as a function of position. The
extreme position is now at $x=0$, and since the plot is symmetric, only the negative part is shown. As is seen, the major part of the total energy remains in the first few sidebands. To make this explicit, also is plotted the remainder energy, which is the energy in the sidebands not shown in this picture.

\section{Down-stream evolution and generation} \label{dseg}

As mentioned in the introduction, part of the motivation for this research is the practical problem to generate large waves in hydrodynamic laboratories. 
In this section, we will show how the previous information can be exploited for this practical purpose. At the same time, it gives a framework to look at the
plots, amplification aspects, and the concept of the MTA.

\subsection{Characteristic down-stream evolutions} \vspace*{0.25cm}

In interpreting the graphical illustrations of spatial evolution of $\eta_\text{SFB}(x,t)$ it is helpful to use the maximal amplitude $M$ and the maximal amplitude amplification $\alpha = M/\left(2r_{0} \right)$. Actually, instead of the original basic variables $\left(\omega, \, r_{0}, \, \hat{\nu}\right)$, one could also use $\left(\omega, \, M, \, \hat{\nu}\right)$ or $\left(\omega, \, M, \, \alpha\right)$ as the parameters of the family.

For the purpose of this section, we will use laboratory variables and consider a wave tank of $5$ m depth. The waves are generated by a wave maker at the left at 
$x = 0$, and travel to the right. We assume the tank is at least $150$ m long and that reflections at an artificial beach can be neglected. The gravitational acceleration is taken to be $9.8$ m/s$^{2}$.

For the three parameters of SFB, the following choices are made. The monochromatic frequency is taken to be $\omega = 3.5$~s$^{-1}$; this will lead to waves with the wavelength of approximately the depth of the layer, so relatively short waves. For the maximum amplitude at the extreme position, we take $M = 0.5$~m. This would correspond to waves of maximal wave height approximately $0.8$--$0.9$~m, which can be considered to be `extreme' waves. For the remaining variable the normalized frequency $\hat{\nu}$ is taken, and we will consider three cases: $\hat{\nu} = \sqrt{3/2}$, $1$ and $\sqrt{1/2}$. For illustrative purposes, the chosen values are to some extent arbitrary, but the chosen ones have some characteristic property: $\hat{\nu} = \sqrt{3/2}$ is the largest value till where phase singularities are present, for $\hat{\nu} = 1$ the B-F growth rate is maximal, while for $\hat{\nu} = \sqrt{1/2}$ the original center frequency of the monochromatic in the B-F signal in the far field vanishes at the extreme position.

For the three cases to be considered, the following table summarizes the main values of the physical quantities.
\begin{table}[ht]
\caption{Physical values of relevant quantities (in m, sec) for three cases of $\hat{\nu}$.} \label{table1}
\begin{center}{\renewcommand{\arraystretch}{1.05}
\begin{tabularx}{0.99\textwidth}{@{}X*{8}{R}@{}}
\toprule
$\quad \hat{\nu}$ & $\nu \quad$ & $\alpha \quad$ & $2r_{0} \quad$ & $T_{\text{mod}} \; \;$ & $N_{\text{temp}}$ & $\zeta \;\;$ & $\lambda_{\text{mod}} \quad$ & $N_{\text{spat}}$ \\ \hline
$\sqrt{3/2}$ & $0.6238$ & $2.0000$ & $0.2500$ & $10.0723$ & $\:\:5.61$ & $5.04$ & $14.1024$ & $2.81$ \\
$1$ & $0.4220$ & $2.4142$ & $0.2071$ & $14.8906$ & $\:\:8.29$ & $3.72$ & $20.8466$ & $4.15$  \\
$\sqrt{1/2}$ & $0.2636$ & $2.7321$ & $0.1830$ & $23.8337$ & $13.28$ & $3.62$ & $33.3700$ & $6.64$ \\
\bottomrule
\end{tabularx}
}
\end{center}
{\footnotesize The maximal amplitude is fixed to be $M = 0.5$~m, and the frequency is fixed to be $\omega=3.5$~s$^{-1}$. 
The frequency corresponds to a temporal period of $T = 1.7952$~s, wavenumber $k=1.25$~m$^{-1}$ and wavelength $\lambda = 5.0265$~m. 
In the successive columns are listed the value of: the normalized modulation frequency $\hat{\nu}$, the non-normalized modulation frequency $\nu$, the amplitude
amplification factor $\alpha$, the asymptotic level of the background $2r_{0}$, the temporal modulation period $T_{\text{mod}} = 2\pi/\nu$, the number of waves in one temporal modulation period defined as $N_{\text{temp}} := T_{\text{mod}}/T$, the time $\zeta$ of the phase singularity (when the largest wave is at $t = 0$), 
the modulation wavelength $\lambda_{\text{mod}}$, and the number of spatial waves in one modulation: $N_{\text{spat}} := \lambda_{\text{mod}}/\lambda$. \par}
\end{table}

Note that a given perturbation $\nu$ in frequency corresponds to a perturbation $\kappa$ in wavenumber according to the dispersion relation:
$\omega + \nu = \Omega(k + \kappa) \approx \Omega(k) + \Omega^{\prime}(k)\kappa$, hence approximately $\nu = V(k)\kappa$. 
The spatial modulation length $\lambda_{\text{mod}}$ can then be approximated by $2\pi/\kappa$ giving $\lambda_{\text{mod}} = V(k) \ast T_{\text{mod}}$. 
Furthermore $N_{\text{temp}} = \omega/\nu \approx \Omega(k)/\Omega^{\prime}(k) \kappa \approx 2k/\kappa$ and so the number of waves in one temporal envelope is approximately twice the number of waves in one spatial envelope.

\begin{figure}[h]
\begin{center}
\includegraphics[width = 0.35\textwidth]{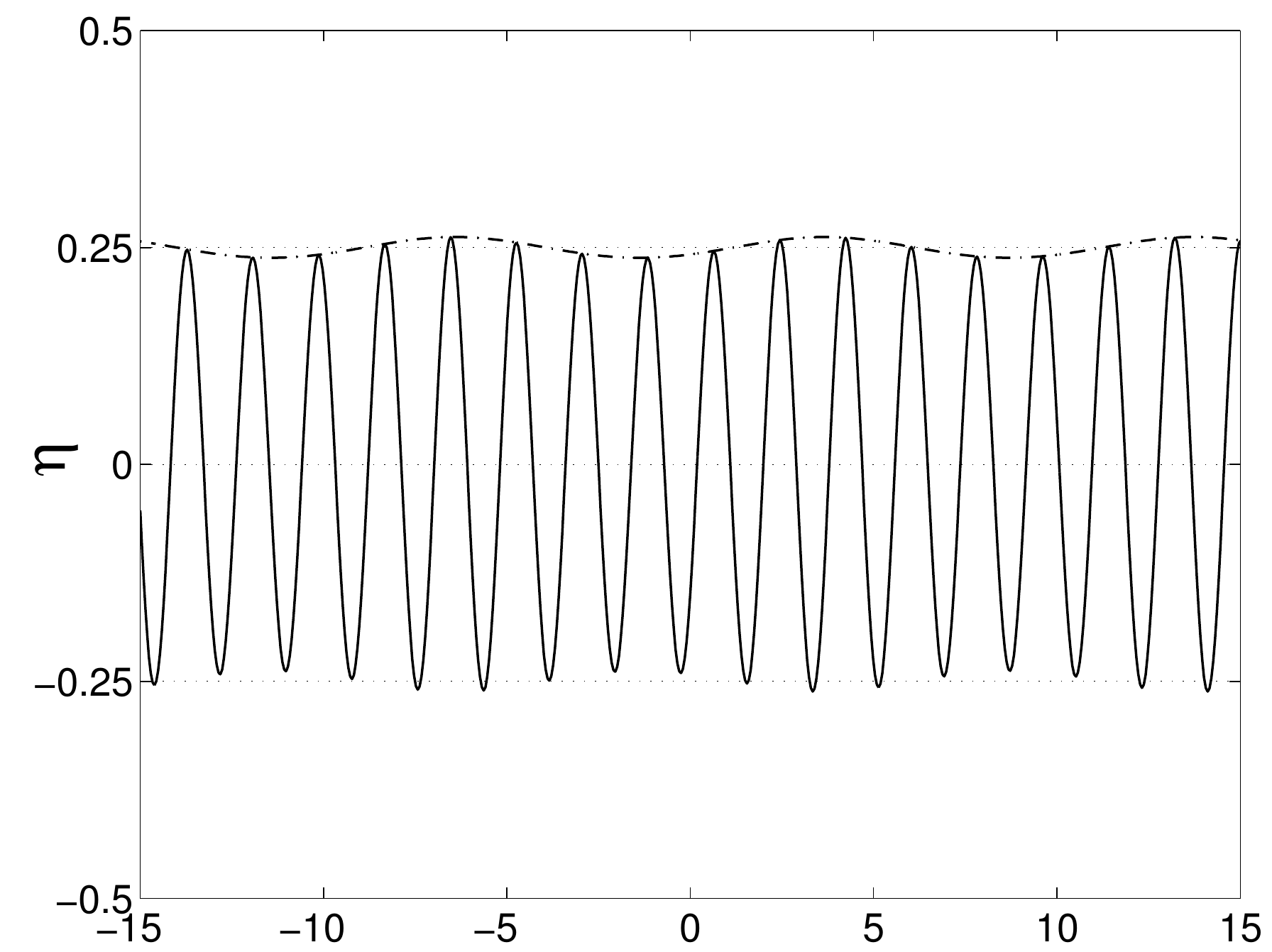} \qquad \qquad 
\includegraphics[width = 0.35\textwidth]{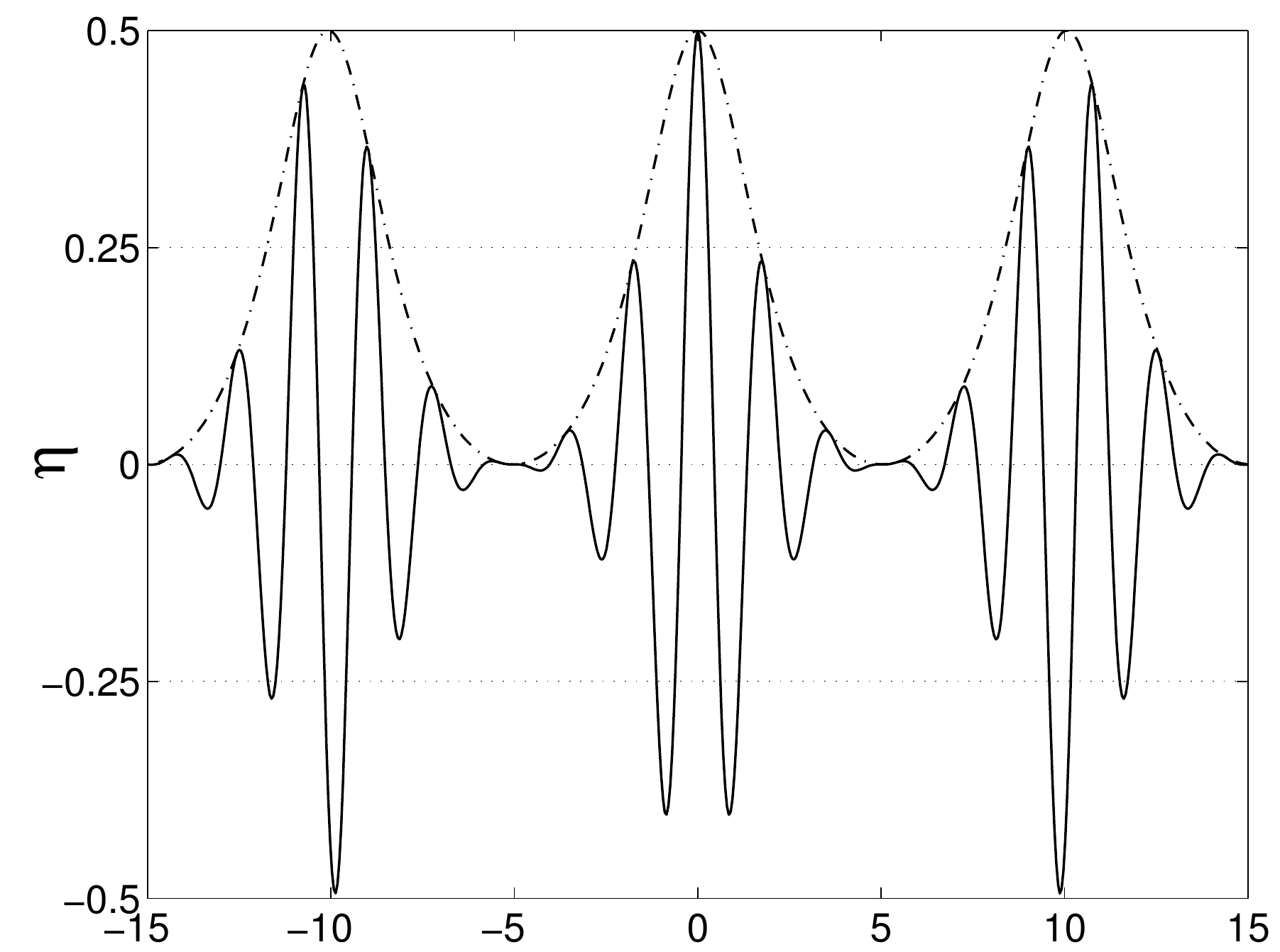} \vspace*{0.25cm}
\includegraphics[width = 0.35\textwidth]{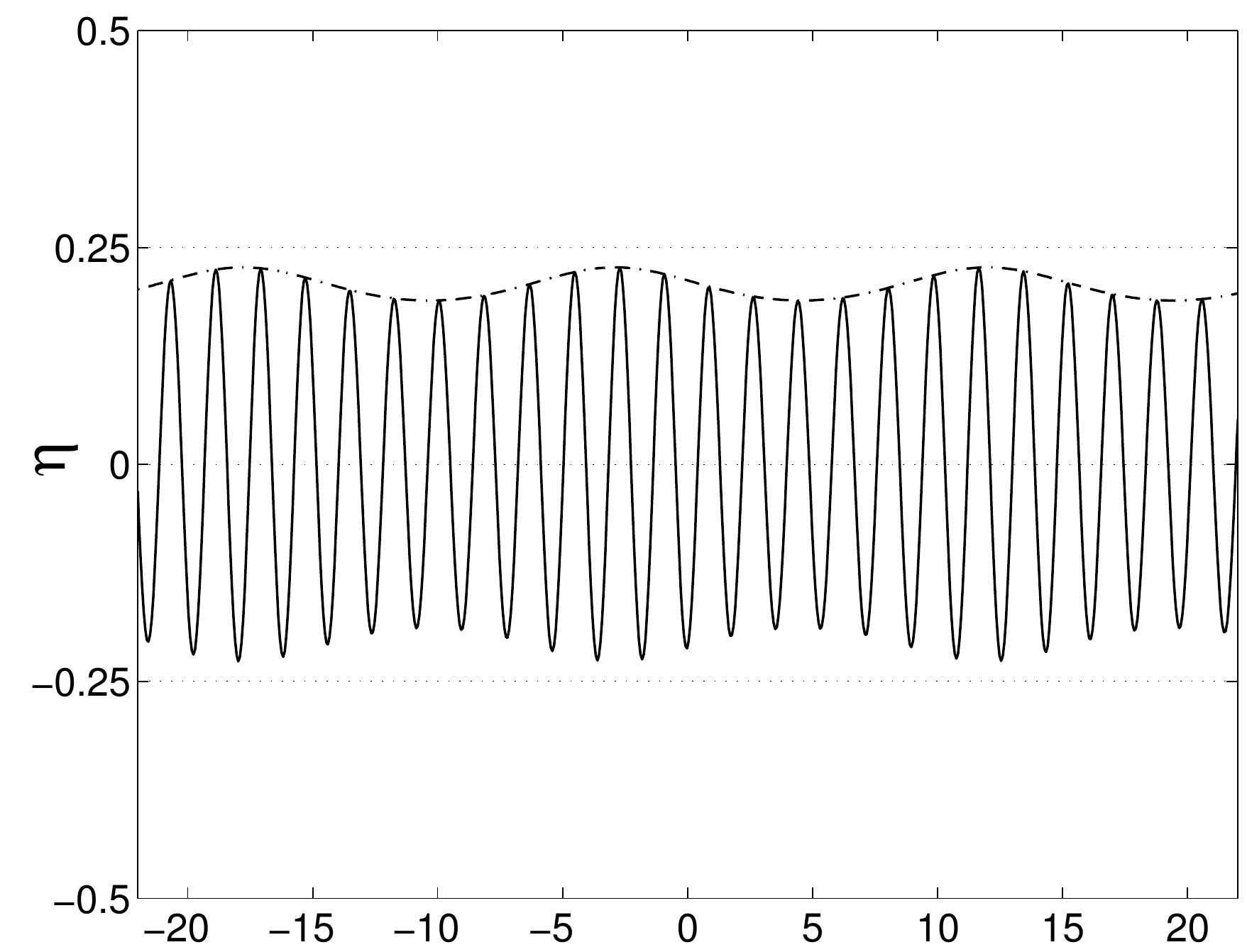} \qquad \qquad 
\includegraphics[width = 0.35\textwidth]{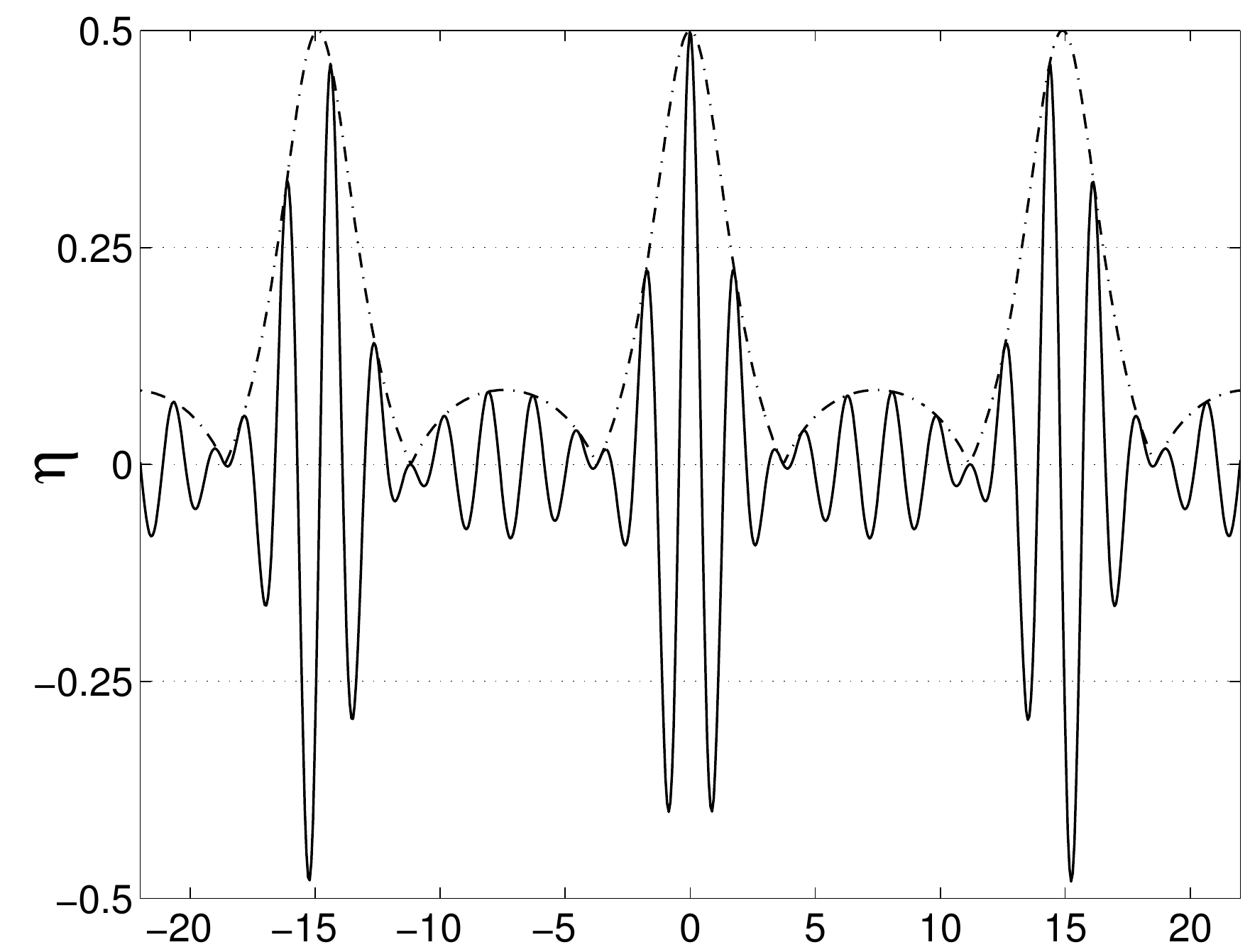} \vspace*{0.25cm}
\includegraphics[width = 0.35\textwidth]{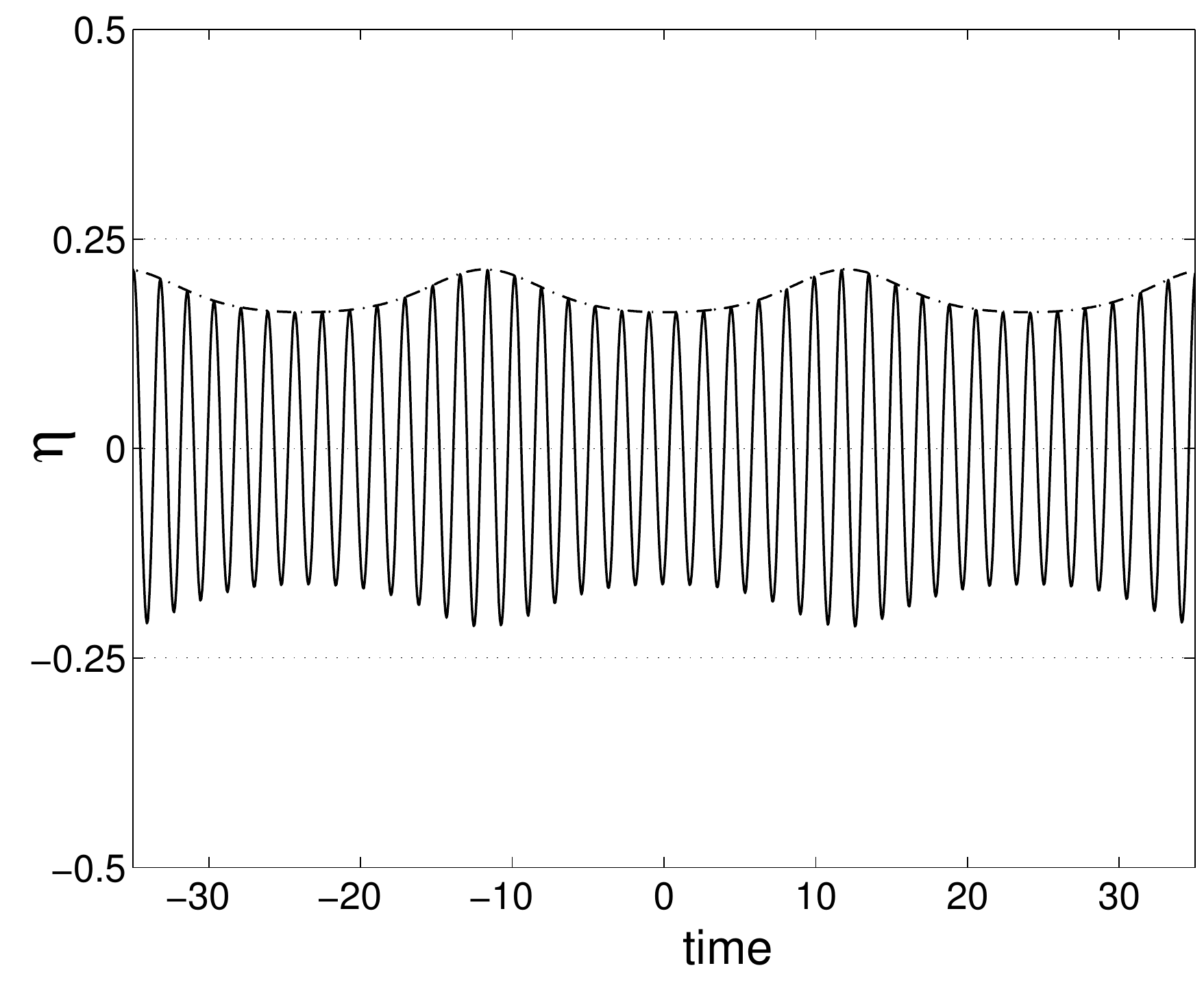} \qquad \qquad 
\includegraphics[width = 0.35\textwidth]{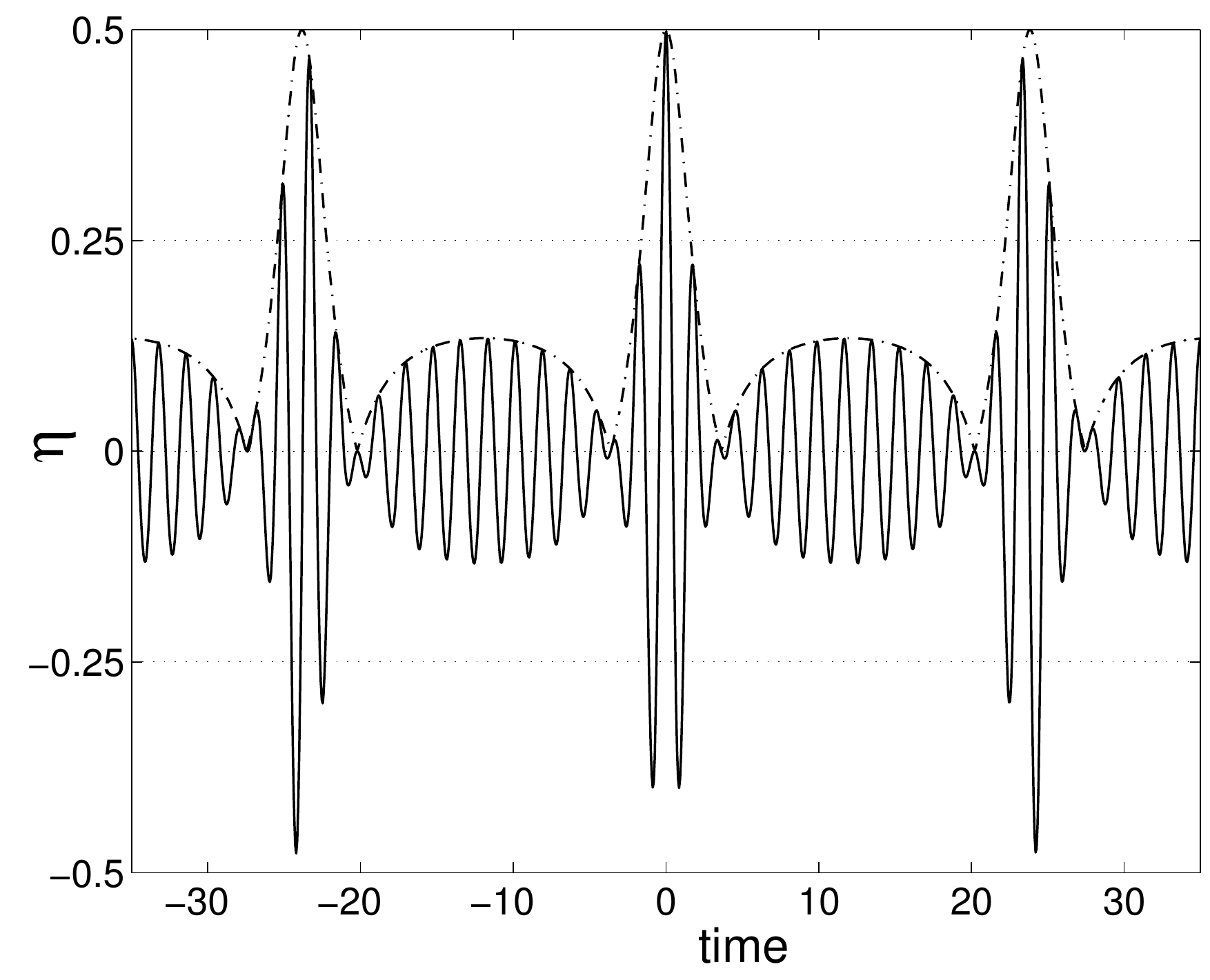} 
\end{center}
\vspace*{-0.5cm} \caption{Plots of time signals of the wave elevation vs time (note the different scaling in time in different rows) at two positions: in the second column at the point of maximal elevation (the extremal signal), and in the first column at a position $150$ m in front of the maximal position. The three rows correspond to different choices for the modulation: from top down $\hat{\nu} = \sqrt{3/2}$, $1$ and $\sqrt{1/2}$.} \label{12}
\end{figure}

In Fig.~\ref{12} samples of the signals at the extreme position as well as at a distance $150$~m are presented for the three cases.
The signal at the distance $150$ m can be interpreted as the signal at the wave maker at $x=0,$ that evolves into the extreme signal at $x = x_{\max} = 150$ m.

Some general observations can be made about the plots. Of course, the large deformations in the signal profiles are clearly visible. Apart from the amplitude amplification $\alpha$ listed in the table for these three cases, a more practical amplification factor would now be the quotient of the maximal wave amplitudes at the two positions. So in this set up $\alpha_{\text{pract}} = \alpha(x_{\max})/\alpha(0)$ will give the amplification in the interval of $150$ m; the numerical values are $\alpha_{\text{pract}} = 1.91$, $2.20$, and $2.34$ for $\hat{\nu} = \sqrt{3/2}$, $1$ and $\sqrt{1/2}$, respectively. Concerning the wave profiles, there is a
clear difference between the three cases. For $\hat{\nu} = \sqrt{3/2}$ the complete signal seems to consist of (relatively `mild') extreme waves, and intermittent waves are missing. That is understandable since this is the limiting case at which the phase singularity start to develop. In the other cases, the number of intermittent waves increases for smaller modulation.

\subsection{MTA} \vspace*{0.25cm}

The maximum temporal amplitude, MTA, has already been defined as
\begin{equation}
\mu(x) = \max_{t}\eta(x,t) \label{MTAdefinition}
\end{equation}
(see \cite{AB}), where $\eta(x,t)$ is the wave elevation. This quantity is of much theoretical and practical importance. By its definition, it describes the largest wave amplitude that can appear at a certain position. On the other hand, it can be seen as a kind of stationary `envelope' of the wave group envelopes: it
describes the boundary where the wall of the wave tank is wet after a long time of waves running downstream.\footnote{We thank Gert Klopman for this illustrative description.} For the special SFB-case under consideration, the MTA can be calculated using the explicit expressions. However, in general, it is not clear how to
obtain theoretically an expression for this function from the nonlinear dynamical equations itself; the dependence on the initial signal will be crucial. Some preliminary investigations will be reported elsewhere.
\begin{figure}[h]
\vspace*{-0.75cm}
\begin{center}
\includegraphics[width = 0.8\textwidth]{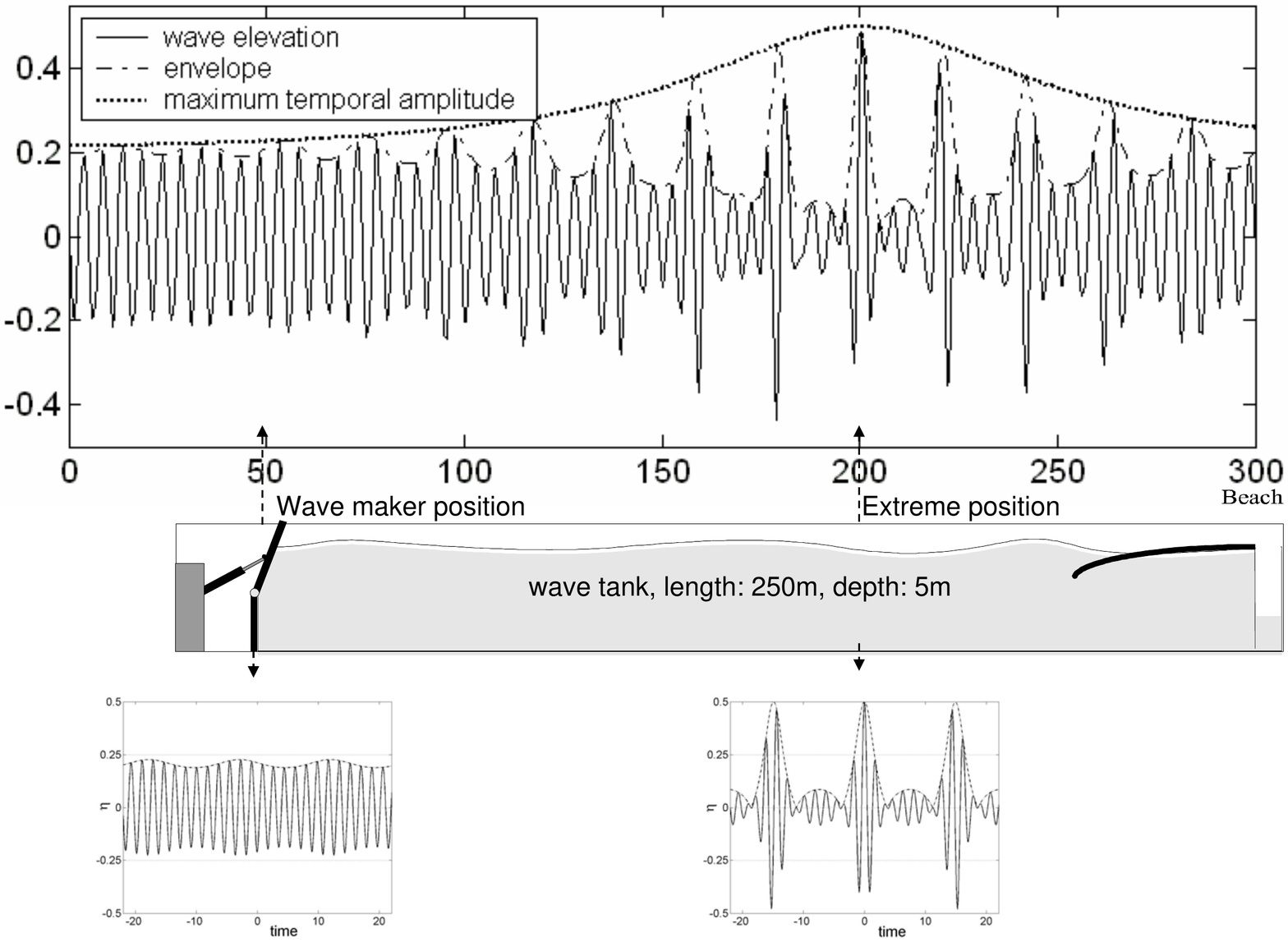}
\end{center}
\vspace*{-1cm} \caption{From knowledge of the MTA of a certain initial signal, the extreme position can be taken to be positioned at a desired place in the wave tank. The distance to the wavemaker then determines which signal has to be taken as the generating initial signal at the wavemaker. The plots are made with the SFB solution for $\hat{\nu} = 1$.} \label{14}
\end{figure}

\begin{figure}[h]
\begin{center}
\includegraphics[width = 0.45\textwidth]{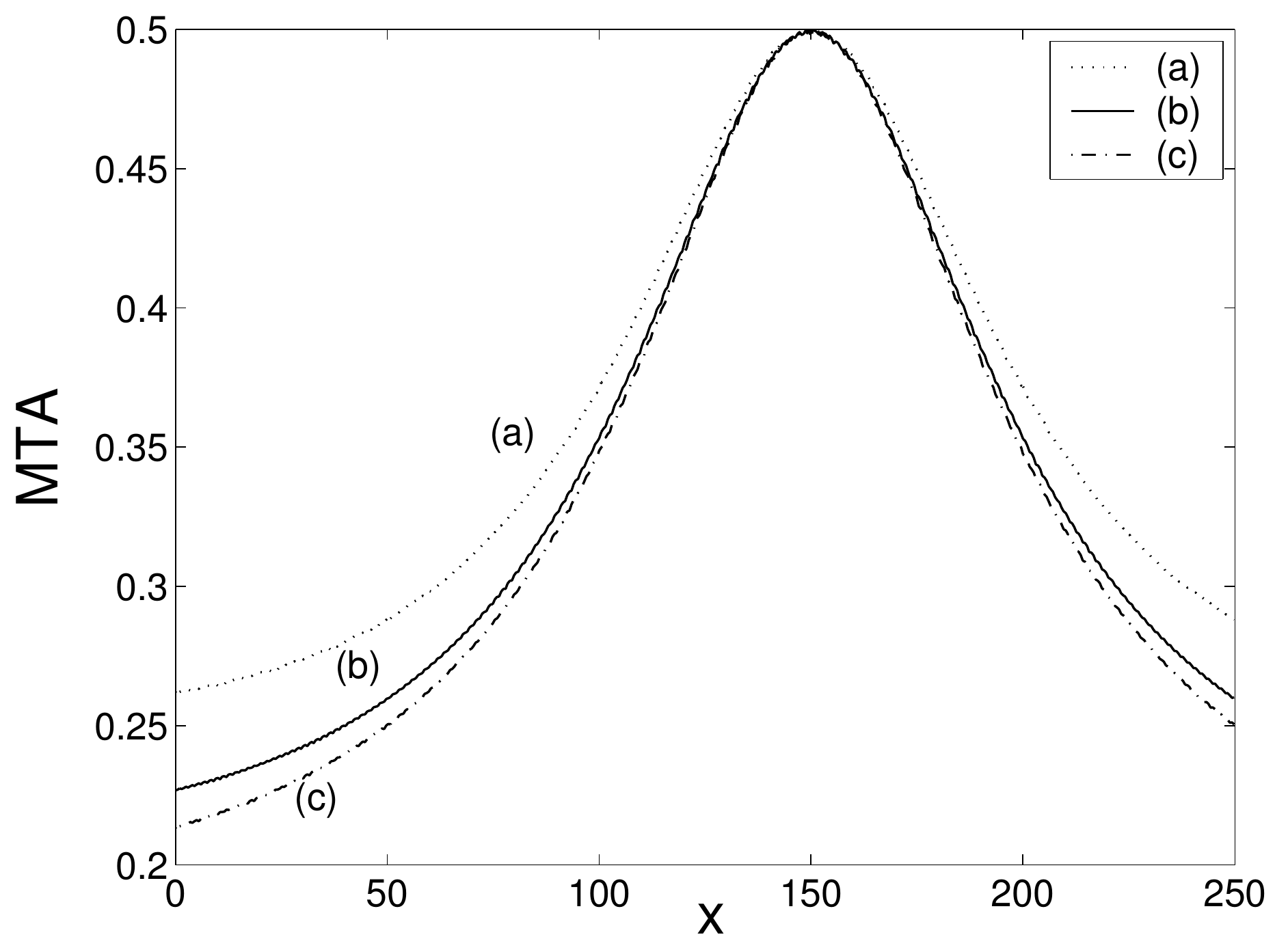}
\end{center}
\vspace*{-0.5cm}
\caption{Plots of the maximal temporal amplitude MTA for three values $\hat{\nu}=\sqrt{3/2}$, $1$ and $\sqrt{1/2}$.}
\label{15}
\end{figure}

From a practical point of view, the MTA is very useful in the design of a generation strategy. For an initial signal to be generated by the wave maker, the MTA will depict the point, say $x_{\max}$, of maximal possible amplitude: $M = \max_{x}\mu\left(x\right) = \mu\left(x_{\max}\right)$. If this position is outside the wave tank, or too close to the beach, the inverse problem could (theoretically) be used: if $L$ is the distance from the wavemaker where the maximal waves are desired to occur, the signal at the position $x_{\max} - L$, i.e. $t\rightarrow\eta\left(x_{\max} - L, t\right)$, should be taken as initial signal at the wavemaker (provided the technical-physical constraints of the wave maker -- such as too high amplitudes, or too large frequencies -- do not prevent this). The practical amplification
factor that will be achieved is then $M/\mu(x_{\max} - L)$. In general, a direct calculation of the inverse problem will be difficult, but in the case of SFB evolution, this is explicit. For SFB evolution we illustrate the above-described methodology in Fig.~\ref{14}, using arbitrarily the value $\hat{\nu}=1$.
For the cases considered in Table~\ref{table1}, the MTA curves are plotted in Fig.~\ref{15}. From this figure the practical amplification factors given earlier can be
read off: $\alpha_{\text{pract}} = \mu(150)/\mu\left(0 \right)$. An expression for the MTA can be found from the explicit expression (2.2); after some simplifications it is found to be given by
\begin{equation}
\left(\frac{\mu (x)}{2r_{0}}\right)^{2} = 1 + \frac{2\hat{\nu}^{2} \sqrt{1 - \hat{\nu}^{2}/2}}{\cosh(\sigma x) - \sqrt{1 - \hat{\nu}^{2}/2}}.  \label{MTAexplicit}
\end{equation}
Observe in this expression the dependence on $\hat{\nu}$, and hence on the initial signal.

\section{Summary and concluding remarks}

In order to understand various aspects of the nonlinear distortion of waves running downstream, we have investigated a family of specific explicit solutions based on the approximate model described by the NLS equation. This family of solitons on finite background, SFB, describe the full nonlinear continuation of the linear modulational, Benjamin-Feir, instability. The exponential growth of the linear instability of a sufficiently long modulated monochromatic wave, is bounded by the nonlinearity, and returns to its original monochromatic character after a long distance. During the downstream evolution, however, large amplitude amplification and wave distortions take place. Such properties are often observed, but difficult to describe from the full nonlinear governing equations. The explicit expressions that are available for SFB have been used to study in detail various aspects that are related to the distortions. These explicit descriptions may contribute to understand `extreme' waves also in more general situations.

Depending on the origin of the physical equation, the practical interpretation of the described qualitative phenomena and quantitative results may differ. In this paper we have concentrated on the behaviour of gravity driven surface waves, and its applicability to generate in a deterministic way extreme waves in a hydrodynamic laboratory. A ship positioned at the extreme position will experience waves described by the extreme signal described above: the intermittent waves, which are numerous when $\hat{\nu}$ is small, resemble a very calm `sea' and are then followed by one or a few large waves (of height that can be larger than three times the wave height in the calm sea), and a quick return to the calm sea, which is repeated periodically in time. This resembles somewhat as if a `freak', or `rogue' wave suddenly appears, as suggested in \cite{Osb00}.

\section*{Acknowledgement}
The authors are very grateful to Dr Pearu Peterson, Ir Gert Klopman and Dr Ren\'{e} Huijsmans for fruitful discussions about this research. This research is conducted partly at Institut Teknologi Bandung, Indonesia and partly at the University of Twente, The Netherlands, and is supported by Riset Unggulan Terpadu Internasional (RUTI 2002/2005) of the Indonesian Ministry of Research and Technological Development, project `On the onset of breaking waves', and by the Netherlands Organisation of Scientific Research NWO, subdivision Applied Sciences STW, project `Extreme Waves' TWI.5374. Partial support from Hibah Bersaing 2002/2004 is also acknowledged.


\begin{thebibliography}{99}
{\footnotesize
\bibitem{AkhAnk} N.N. Akhmediev and A. Ankiewicz, Solitons--Nonlinear pulses and beams, Chapman \& Hall, London, 1997.

\bibitem{Whitham} G.B. Whitham, Linear and Nonlinear Waves, John Wiley \& Sons, New York, 1974.

\bibitem{JHGrRene01} J.H. Westhuis, E. van Groesen, R.H.M. Huijsmans, Long time evolution of unstable bichromatic waves, in: T. Miloh (Ed.), 
15th International Workshop on Water Waves and Floating Bodies, Tel Aviv, Israel, 2000, pp.~184--187. ISBN 965-274-288-0.

\bibitem{JHGrRene02} J.H. Westhuis, E. van Groesen, R.H.M. Huijsmans, Experiments and numerics of bichromatic wave groups, 
J. Waterway, Port, Coast. Ocean Eng. 127 (2001) 334--342.

\bibitem{GrJH01} E. van Groesen, J.H. Westhuis, Modelling and simulation of surface water waves, Math. Comput. Simulat. 59 (2002) 341--360.

\bibitem{JH} J.H. Westhuis, The numerical simulation of nonlinear waves in a hydrodynamic model test basin, PhD thesis, University of Twente, The Netherlands, 2001.

\bibitem{JHAan} J.H. Westhuis, Andonowati, Applying the Finite Element Method in numerically solving the two dimensional free-surface water wave equations,
in: A.J. Hermans (Ed.), Proceedings of the 13th International Workshop on Water Waves and Floating Bodies, Alphen a/d Rijn, The Netherlands, 1998,
pp.~171--174.

\bibitem{JH02} J.H. Westhuis, Approximate analytic solutions and numerical wave tank results for the reflection coefficients of a class of numerical beaches,
in: Proceedings of the 10th ISOPE Conference, Seattle, Washington, 3, 2000, pp.~242--252. ISBN 1-880653-49-4.

\bibitem{Dean} R.G. Dean, Freak waves: a possible explanation, in: A. T{\o}rum, O.T. Gudmestad (Eds.), Water Wave Kinematics, 
Kluwer Academic Publishers, Dordrecht, 1990, pp.~609--612.

\bibitem{Donelan} M.A. Donelan, W.H. Hui, Mechanics of ocean surface waves, in: G.L. Geernaert, W.J. Plants (Eds.), Surface Waves and Fluxes, vol. 1, 
Kluwer Academic Publishers, Dordrecht, 1990, pp.~209--246.

\bibitem{GroAanKar04Brest} E. van Groesen, Andonowati, N. Karjanto, Deterministic aspects of nonlinear modulation instability, 
in: M. Olagnon, M. Prevosto (Eds.), Proceedings Rogue Waves 2004, Brest France, IFREMER, 2005, 12~pp. ISBN 2-84433-150-5.

\bibitem{Longuet} M.S. Longuet-Higgins, Statistical properties of wave groups in a random sea state, Philos. Trans. Roy. Soc. London, A312 (1984) 219--250.

\bibitem{Osb00} A.R. Osborne, M. Onorato, M. Serio, The nonlinear dynamics of rogue waves and holes in deep-water gravity wave trains, 
Phys. Lett. A 275 (2000) 386--393.

\bibitem{Phillips} O.M. Phillips, D.Gu, M.A. Donelan, Expected structure of extreme waves in a Gaussian sea. Part I. Theory and SWADE buoy measurements, 
J. Phys. Oceanogr. 23 (1993) 992--1000.

\bibitem{Dysthe} K. Dysthe, Note on a modification to the nonlinear Schr\"{o}dinger equation for application to deep water waves, 
Proc. R. Soc. London, Ser. A 369 (1979) 105--114.

\bibitem{Hen99} K.L. Henderson, D.H. Peregrine, J.W. Dold, Unsteady water wave modulation: fully nonlinear solutions and comparison with the nonlinear Schr\"{o}dinger equation, Wave Motion 29 (1999) 341--361.

\bibitem{BenjaminFeir} T.B. Benjamin, J.E. Feir, The disintegration of wave trains on deep water. Part 1. Theory, J. Fluid Mech. 27 (1967) 417--430.

\bibitem{AB} Andonowati, E. van Groesen, Optical pulse deformation in second order non-linear media, J. Nonlinear Opt. Phys. Mater. 12 (2003) 221--234.

\bibitem{AanRnBr} Andonowati, W.M. Kusumawinahyu, E. van Groesen, Steepness of extreme waves caused by wave focussing in the Benjamin-Feir instability regime, 
in: G.H. Jirka, W.S.J. Uijttewaal (Eds.), Shallow Flows: Research Presented at the International Symposium on Shallow Flows, Delft, The Netherlands, June 16--18, 2003, pp.~205--209, CRC Press, Taylor \& Francis Group, Boca Raton, Florida, 2004.

\bibitem{AanRnBrJPO} Andonowati, W.M. Kusumawinahyu, E. van Groesen, A numerical study of the breaking of modulated waves generated at a wavemaker, 
Appl. Ocean Res. 28 (2006) 9--17.

\bibitem{Osb01} M. Onorato, A.R. Osborne, M. Serio, S. Bertone, Freak waves in random oceanic sea states, Phys. Rev. Lett, 86 (2001) 5831.

\bibitem{Akhmediev87} N.N. Akhmediev, V.M. Eleonskii, N.E. Kulagin, Exact first-order solutions of the nonlinear Schr\"{o}dinger equation,
Theoret. Math. Phys. 72 (1987) 809--818.

\bibitem{Dysthe99} K. Dysthe, K. Trulsen, Note on breather type solutions of the NLS as models for freak-waves, Phys. Scripta T82 (1999) 48--52.

\bibitem{Karjanto04} N. Karjanto, E. van Groesen, Derivation of the NLS breather solutions using displaced phase-amplitude variables,
in: Proceedings of the 5th SEAMS-GMU Conference 2007, 2008, pp.~357--368. 

\bibitem{Ma79} Y.-C. Ma, The perturbed plane-wave solutions of the cubic Schr\"{o}dinger equation, Stud. Appl. Math. 60 (1979) 43--58.

\bibitem{Peregrine83} D.H. Peregrine, Water waves, nonlinear Schr\"{o}dinger equations and their solutions, J. Aust. Math. Soc. Ser. B 25 (1983) 16--43.

\bibitem{GroAanKar06DPA} E. van Groesen, Andonowati, N. Karjanto, Displaced phase-amplitude variables for waves on finite background, Phys. Lett. A 354 (2006) 312--319.

\bibitem{Ert00} R.C. Ertekin, J.W. Kim, The study of the spatial coherence of surface waves by the nonlinear Green-Naghdi model in deep water, 
Annual Report No.~ADA609930, Hawaii University Honolulu (Manoa Campus), Defense Technical Information Center, US Department of Defense, 2000, pp.~1--8.
Available online at \url{http://www.dtic.mil/get-tr-doc/pdf?AD=ADA609930}.

\bibitem{Zak02} E. Zakharov, A.I. Dyachenko, O.A. Vasilyev, New method for numerical simulation of a nonstationary potential flow of incompressible fluid with a free surface, Eur. J. Mech. B/Fluids 21 (2002) 283--291.

\bibitem{KarGroPet02} N. Karjanto, E. van Groesen, P. Peterson, Investigation of the maximum amplitude increase from the Benjamin-Feir instability, 
J. Indones. Math. Soc. (MIHMI) 8 (2002) 39--47.

\bibitem{expMARIN04} R. Huijsmans, G. Klopman, N. Karjanto, Andonowati, Experiments on extreme wave generation using the Soliton on Finite Background, 
in: M. Olagnon, M. Prevosto (Eds.), Proceedings Rogue Waves 2004, Brest France, IFREMER, 2005, 10~pp. ISBN 2-84433-150-5.

\bibitem{Gro98} E. van Groesen, Wave groups in uni-directional surface wave models, J. Eng.Math. 34 (1998) 215--226.

\bibitem{Tan95} M. Tanaka, Disappearance of waves in modulated train of surface gravity waves, in: A. Mielke and K. Kirchg\"{a}ssner (Eds.), Proceedings of IUTAM/SIMM Symposium on Structure and Dynamics of Nonlinear Waves in Fluids, Advanced Series in Nonlinear Dynamics, vol. 7, World Scientific, Singapore, 1995, pp. 392--398.

\bibitem{Dold92} J.W. Dold, An efficient surface-integral algorithm applied to unsteady gravity waves, J. Comput. Phys. 103 (1992) 90--115.

\bibitem{Cos99} C. Coste, F. Lund, M. Umeki, Scattering of dislocated wave fronts by vertical vorticity and the Aharonov-Bohm effect.
I. Shallow water, Phys. Rev. E 60 (1999) 4908--4916.

\bibitem{BalEA00} M.L.M. Balistreri, J.P. Korterik, L. Kuipers and N.F. van Hulst, Local observations of phase singularities in optical fields in waveguide structures, Phys. Rev. Lett. 85 (2000) 294.

\bibitem{Bas95} I.V. Basistiy, M.S. Soskin, M.V. Vasnetsov, Optical wavefront dislocations and their properties, Opt. Comm. 119 (1995) 604--612.

\bibitem{Fre99} I. Freund, Critical point explosions in two-dimensional wave fields, Opt. Comm. 159 (1999) 99--117.

\bibitem{Gbur02} G. Gbur, T.D. Visser, E. Wolf, Anomalous behavior of spectra near phase singularities of focused waves, Phys. Rev. Lett. 88 (2002) 013901.

\bibitem{NyeBer74} J.F. Nye, M.V. Berry, Dislocation in wave trains, Proc. R. Soc. London Ser. A 336 (1974) 165--190.

\bibitem{PopDog02} G. Popescu, A. Dogariu, Spectral anomalies at wave-front dislocations, Phys. Rev. Lett. 88 (2002) 183902.












}
\end{thebibliography}
\end{document}